\documentclass[article, a4paper]{IEEEtran}
\usepackage{graphicx}
\usepackage{units}
\usepackage{array}
\usepackage{multirow}
 
\usepackage{multirow}
\usepackage{rotating}
\usepackage{textcomp}
\usepackage[cmex10]{amsmath}
\usepackage{enumitem}
\usepackage{color}
\usepackage{soul}

\newcommand{\ACI}{\mathrm{A}}
\newcommand{\mes}{\emph{m}}
\newcommand{\msg}{\emph{M}}
\newcommand{\APone}{{\emph{1AP}}}
\newcommand{\APtwo}{{\emph{2APs}}}

\usepackage{hyperref}

\usepackage{lipsum}
\usepackage[pscoord]{eso-pic}

\newcommand{\placetextbox}[3]{
  \setbox0=\hbox{#3}
  \AddToShipoutPictureFG*{
    \put(\LenToUnit{#1\paperwidth},\LenToUnit{#2\paperheight}){\vtop{{\null}\makebox[0pt][c]{#3}}}%
  }%
}%

\begin{document}
\placetextbox{0.5}{1}{This is the author's version of an article that has been published in this journal.}
\placetextbox{0.5}{0.985}{Changes were made to this version by the publisher prior to publication.}
\placetextbox{0.5}{0.97}{The final version of record is available at \href{https://doi.org/10.1109/TII.2017.2759788}{https://doi.org/10.1109/TII.2017.2759788}}%
\placetextbox{0.5}{0.05}{Copyright (c) 2018 IEEE. Personal use is permitted.}
\placetextbox{0.5}{0.035}{For any other purposes, permission must be obtained from the IEEE by emailing pubs-permissions@ieee.org.}%

\title{Improving Effectiveness of Seamless Redundancy in Real Industrial Wi-Fi Networks
\thanks{This work was partially supported by Regione Piemonte and the Ministry of Education, University, and Research of Italy (MIUR) in the framework of the Call ``Fabbrica Intelligente'', Project HuManS ``Human centered Manufacturing Systems'' (application number 312-36). Copyright (c) 2017 IEEE. Personal use of this material is permitted. However, permission to use this material for any other purposes must be obtained from the IEEE by sending a request to pubs-permissions@ieee.org. The authors are with the National Research Council of Italy, Istituto di Elettronica e di Ingegneria dell'Informazione e delle Telecomunicazioni (CNR-IEIIT), I-10129 Torino, Italy (e-mail: {name.surname}@ieiit.cnr.it).}}

\author{Gianluca~Cena, \textit{Senior Member}, \textit{IEEE}, Stefano~Scanzio, \textit{Member}, \textit{IEEE}, and\\Adriano~Valenzano, \textit{Senior Member}, \textit{IEEE}}

\maketitle

\begin{abstract}
Reliability and determinism of Wi-Fi can be tangibly improved by means of seamless redundancy, to the point of making this technology suitable for industrial environments.
As pointed out in recent papers, the most benefits can be achieved when no phenomena can simultaneously affect transmissions on all channels of a redundant link.

In this paper several aspects are analyzed which, if not properly counteracted, may worsen seamless redundancy effectiveness.
Effects they cause on communication have been experimentally evaluated in real testbeds, which rely on commercial \mbox{Wi-Fi} devices.
Then, practical guidelines are provided, which aim at preventing joint interference through a careful system design.
Results show that measured communication quality can be made as good as expected in theory.
\end{abstract}

\IEEEpeerreviewmaketitle


\section{Introduction}
The past decade witnessed a dramatic growth of the IEEE 802.11 \cite{2016-std-80211} technology (best known as Wi-Fi) as the standard wireless local area network (WLAN) solution in many application areas, including industrial environments \cite{Willig_2008}.
Notable exceptions are real-time control applications running at the shop-floor, as \mbox{Wi-Fi} is currently deemed unable to satisfy determinism and reliability requirements demanded by time- and safety-critical systems.
For this reason, other wireless solutions like WirelessHART \cite{2011-IEM-Petersen} are mostly being considered in these scenarios.
Nonetheless, a number of proposals have lately appeared which concern the use of Wi-Fi in distributed real-time control systems.
For instance, \cite{2007-TII-vas} and \cite{2011-ISPCS-TDMA-flexWARE} rely on deterministic overlays placed atop the IEEE 802.11 Medium Access Control (MAC) layer, those in \cite{Gamba_et_al_2010, 2017-TII-EDF}, and \cite{2016-TII-Tian} foresee modifications to either the MAC retransmission scheme or the backoff mechanism, while \cite{2010-TII-802.11e} investigates the use of IEEE 802.11e Quality-of-Service (QoS).
Besides the high throughput, the main advantage of Wi-Fi is that it can be interconnected with Ethernet directly at the data-link layer, which means that hybrid wired-wireless networks can be easily set up \cite{2008-IEM-Hybrid}.
Unlike \cite{2016-TIM-EWMA}, packet losses in wired portions of such systems are typically negligible, and upper bounds can  be often enforced on the related transmission delays. 
This is no longer true for Wi-Fi links, which become the weakest point in the network and negatively affect the overall reliability.

As shown in recent literature, the adoption of seamless redundancy at the data-link layer of wireless networks  improves their behavior significantly, under the basic assumption that disturbance and interference on physical channels are statistically independent.
In \cite{2016-SGComm} and \cite{2017-TII-PRP_REDUNDANCY}, such a hypothesis was experimentally checked for redundant links based on real Wi-Fi equipment.
While tangible improvements were achieved in typical operating conditions, channels were found not to be completely independent.
Not surprisingly, this was not due to nearby wireless stations (STA), whose operations were actually uncorrelated, but depended on the design and configuration of the software and hardware modules of \emph{redundant stations} (RSTA) in the testbeds.

In this paper, the most relevant design aspects that may cause joint and mutual interference among channels in a Redundant Basic Service Set (RBSS), i.e., a Wi-Fi network exploiting seamless redundancy, are identified.
Then, suitable guidelines are provided, which can be used when industrial devices, communicating over an RBSS, have to be implemented and deployed, so as to maximize seamless redundancy benefits.
With respect to \cite{2016-ETFA-Guidelines}, which only reported on preliminary and partial results,
this paper offers a comprehensive and structured overview of all main causes of interference.
Effects of the network infrastructure and deeper insights on the ways transmissions on air from nearby antennas may affect each other are also included.
Moreover, a thorough assessment is made about the residual channel dependence, after such phenomena have been tackled by means of above guidelines.

The paper is organized as follows: Section~II provides a quick survey of Wi-Fi seamless redundancy, while Section~III describes the experimental setup we employed for evaluating the quality of communication.
Sections IV~presents a general view of phenomena which may affect transmissions on redundant links, whereas  Sections~V and VI focus on interference that takes place inside devices and on air, respectively.
Performance of a target solution, where all causes of interference have been dealt with properly, is experimentally assessed in Section~VII.
Finally, Section~VIII summarizes practical design guidelines and draws conclusions.

\section{Related work}
A number of solutions aimed at improving fault tolerance in Industrial Ethernet networks are based on seamless redundancy.
For example, in the Parallel Redundancy Protocol (PRP) \cite{2012-std-PRP} two copies of each packet are sent by the originator on a pair of distinct, similar networks at the same time.
The copy reaching the recipient first is retained, while the late copy is discarded.
This approach can be profitably applied to wireless links (in particular, those based on Wi-Fi) in order to improve reliability and timeliness of data exchanges.
In fact, the \emph{packet loss ratio} (PLR) on the redundant link is noticeably lower than on either single physical channel, and also transmission latencies decrease.

Several proposals have appeared in the past few years where seamless redundancy is exploited in specific application scenarios.
In \cite{2015-TIM-PRP_wired_wireless}, a hybrid wired-wireless redundancy scheme was envisaged to improve reliability in electrical substations, while redundant wireless paths were used in \cite{2016-SGComm} for streaming phasor data over Wi-Fi.
The IP parallel redundancy protocol (iPRP), introduced in \cite{2016-TII-iPRP}, applies the same concepts to geographic networks for use with smart grids.
In \cite{2017-TII-survey_sync} seamless redundancy was proposed for clock synchronization protocols in WLANs, in order to both increase robustness against packet losses and mitigate the effects of malicious security attacks.
In \cite{2017-EUROCON-redundancy_railway}, PRP was applied to 4G Long Term Evolution (LTE) cellular networks in the context of railway wireless communications.
Finally, in \cite{2014-WFCS-Mifdaoui} Ultra WideBand (UWB) technology was considered, possibly on redundant channels, as a backup network for avionic systems, in order to reduce cables and weight.

\begin{figure}[t]
  \scriptsize
  \centering
  \includegraphics[width=1.0\columnwidth]{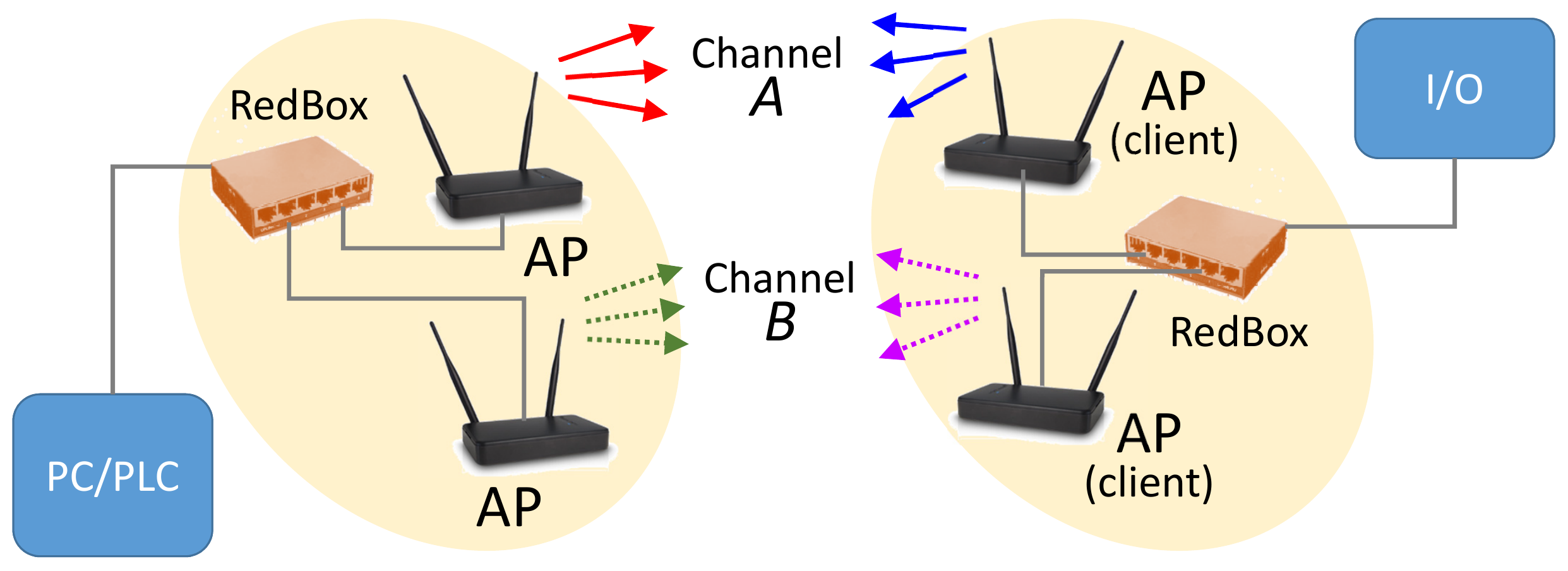}
  \caption{Practical implementation of PRP over Wi-Fi (PoW).}
  \label{fig:pow}
\end{figure}

In \cite{2012-WFCS-WoP1} and \cite{2013-ETFA-Rentschler}, an interesting implementation of seamless redundancy applied to IEEE 802.11, explicitly aimed at factory automation, is described.
Such an arrangement, which is sketched in Fig.~\ref{fig:pow}, relies on commercial off-the-shelf (COTS) Wi-Fi access points (APs), some of which operate in client mode, and special-purpose PRP devices (RedBoxes).
In the following, it will be referred to as \emph{PRP over Wi-Fi} (PoW).
This solution is particularly appealing, as it can be easily employed to provide wireless extensions to industrial networks based on switched Ethernet, like EtherNet/IP and \mbox{SafetyNET p}.
It is worth noting that, in PoW, packet duplication and deduplication at both ends of a redundant wireless link can also be carried out in software, with significantly lower implementation costs.
The \emph{Wi-Fi Redundancy} (Wi-Red) proposal \cite{2016-tii-WiRed} is quite similar to PoW, but it defines specific duplicate avoidance mechanisms, preliminarily introduced in \cite{2014-WFCS-WiRed} and \cite{2014-ETFA-DDD}, which are aimed at reducing network traffic \cite{2017-WFCS-bandwidth} and enhancing the overall system behavior.

As shown in \cite{2016-SGComm} and \cite{2017-TII-PRP_REDUNDANCY}, the most benefits can be obtained from seamless redundancy when phenomena that affect transmissions on distinct physical channels are completely independent.
In this case, in fact, the probability that a packet is lost (or that it misses a certain deadline) on the redundant link can be obtained by multiplying the relevant probabilities, evaluated separately on each physical channel.
For example, if the PLR on each single channel is $1\%$, the overall PLR when redundancy is exploited can be as low as $0.01\%$.

In theory, allocating channels of a redundant link to non-overlapping frequencies is enough to ensure their statistically independent behavior, as sources of interference (nearby STAs) and disturbance (narrowband noise) are uncoordinated.
However, as pointed out in \cite{2016-ETFA-Guidelines}, achieving true independence also requires a number of practical aspects to be taken into account. 
Two main factors may worsen the performance of real redundant Wi-Fi networks: the first one is due to a number of protocol functions, and sometimes depends on the specific implementation of \emph{software} modules of the protocol stack, whereas the second is caused by physical phenomena affecting \emph{hardware} components, including radio transceivers and antennas.
The related interference can occur either \emph{inside} nodes (in both end devices and network equipment) or \emph{on air}.

\section{Experimental testbed}
\label{sec:setup}
\begin{figure}[t]
  \scriptsize
  \centering
  \includegraphics[width=1.0\columnwidth]{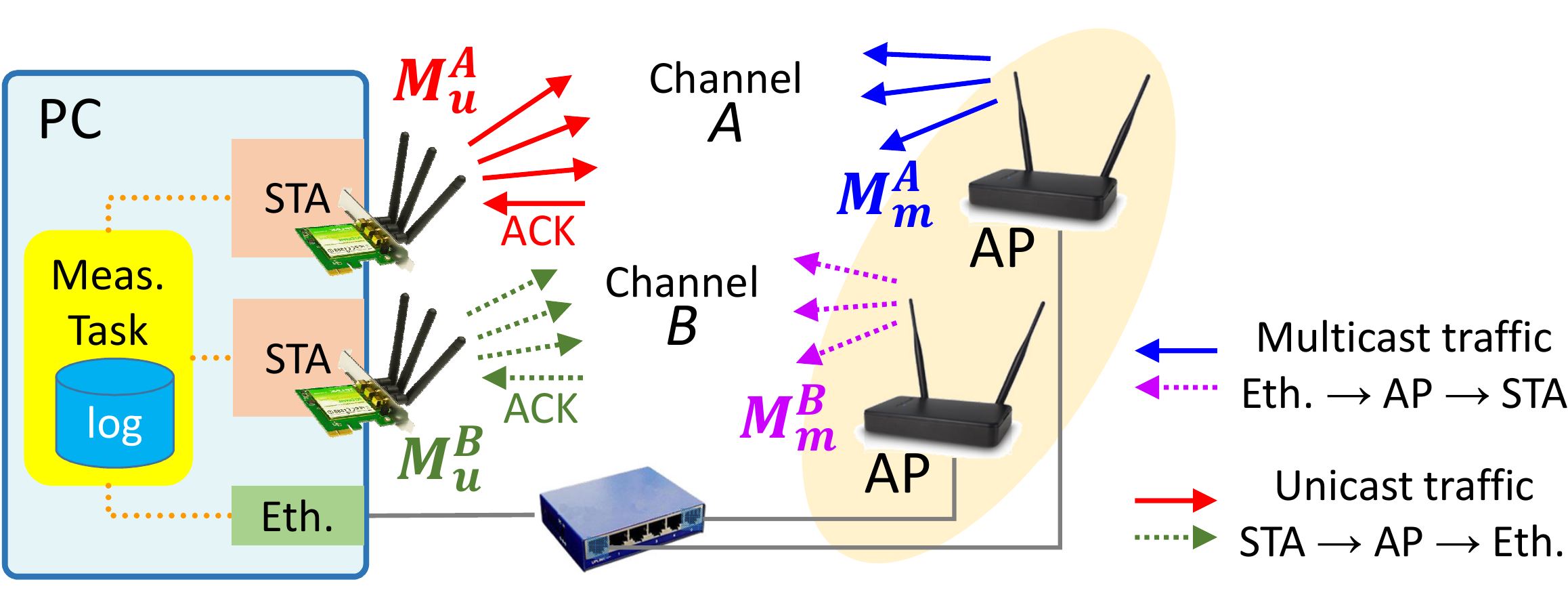}
  \caption{Experimental setup for evaluating communication quality.}
  \label{fig:testbed}
\end{figure}
To investigate what aspects affect the quality of communication on redundant links, a simplified testbed was developed, which retains the same features as full-fledged PoW implementations like the one in \cite{2012-WFCS-WoP1}.
Then, an experimental campaign was carried out and the effects of a number of implementation choices were analyzed.
The aim is drawing a set of practical guidelines for system designers to maximize  benefits achievable with seamless redundancy in Wi-Fi networks.
The setup  sketched in Fig.~\ref{fig:testbed}  is an extension of those described in \cite{2017-TII-PRP_REDUNDANCY} and \cite{2017-TII-EDF}.
Two identical \mbox{Wi-Fi} adapters are installed  in a Linux PC and associated to two identical APs.
The APs are in turn connected to the Gigabit Ethernet port of the PC through a switch.
Equipment of different brands was used in the experiments, so as to ensure that results do not depend on the behavior of a specific device.

A measurement task running on the PC generates a cyclic stream of packets, and transmits them concurrently on the two channels.
The PC acts as both source and destination of packets,
which can be set to travel either from  Wi-Fi adapters to the Ethernet port or in the opposite direction.
Two types of streams can be produced, unicast and multicast: the former is directed from STAs to APs, while the latter flows in the opposite direction.
This is because multicast frames are sent by the originating STA to the AP as unicast.
In the following, subscripts refer to stream types, while superscripts are used to identify channels (channels $1...13$ belong to the $\unit[2.4]{GHz}$ band, while channels $36...165$ lie in the $\unit[5]{GHz}$ band).
For example, $\msg_u^{c11}$ denotes a unicast stream transmitted on channel $11$, whereas $\msg_m^{c165}$ is a multicast stream on channel $165$.
As typical for process data in industrial applications, packets sent by the measurement task were quite small (the payload was set equal to $\unit[50]{B}$).
Unless otherwise specified, and to better adhere to the behavior of real-world networks, we left the rate adaptation mechanism enabled in the experiments.
Therefore, the modulation and coding scheme, and hence the bit rate, were dynamically selected by the originating STA.

By taking timestamps on every single transmission and reception, the latency experienced by each packet (or, in case of errors, an indication that the packet went lost) can be determined on any channel.
Besides transmission on air, latency measured on packet arrival at the Ethernet port, termed \emph{end-to-end latency} ($d$), includes processing times in the PC, forwarding delays in both the AP and the switch, as well as transmission times on Ethernet cables.
With the exception of delays introduced by the AP, which are analyzed in Section \ref{sec:NetworkComponents}, all these contributions were confirmed to be negligible. 
For example, processing times in commercial PCs \cite{2017-WFCS-SDMAC}, which constitute the second largest contribution to delays after the AP, are at least one order of magnitude smaller than the transmission latencies we measured.
In addition, other timestamps were acquired upon reception of the relevant acknowledgment by each wireless adapter.
They include both the interframe space (IFS) and the ACK frame, whose duration can be computed easily.
By subtracting these durations from timestamps taken on ACKs, frame arrival times to the APs can be determined.
In this way, a second latency can be defined for packets originating from wireless adapters, 
termed \emph{link latency} ($d^\prime$), which does not include contributions due to the network infrastructure.

All measures are collected at runtime in the main memory of the PC, saved to disk at the end of each experiment, and processed offline to evaluate the quality of communication.
Starting from data acquired for each physical channel and by applying PRP rules (i.e., considering the copy of each packet which arrives to destination first), relevant performance indexes can be obtained for the redundant link too. 
In practice, transmission latencies $d_i$ and $d^\prime_i$ for both physical channels and the redundant link were computed for every sample $i$ concerning a successfully delivered packet in the experiment log file.
Typical statistical indexes were evaluated for $d_i$, e.g., average value and standard deviation ($\bar d$ and $\sigma_d$), minimum and maximum (worst-case) values ($d_{\operatorname{min}}$ and $d_{\operatorname{max}}$), and percentiles (e.g., $d_{p99.9}$ and $d_{p99.99}$), and for $d^\prime_i$ as well.
Depending on the experiment, statistics on either $d_i$ or $d^\prime_i$ were used.
More details on the testbed can be found in \cite{2017-TII-PRP_REDUNDANCY} and \cite{2017-TII-EDF}.

\begin{figure}[t]
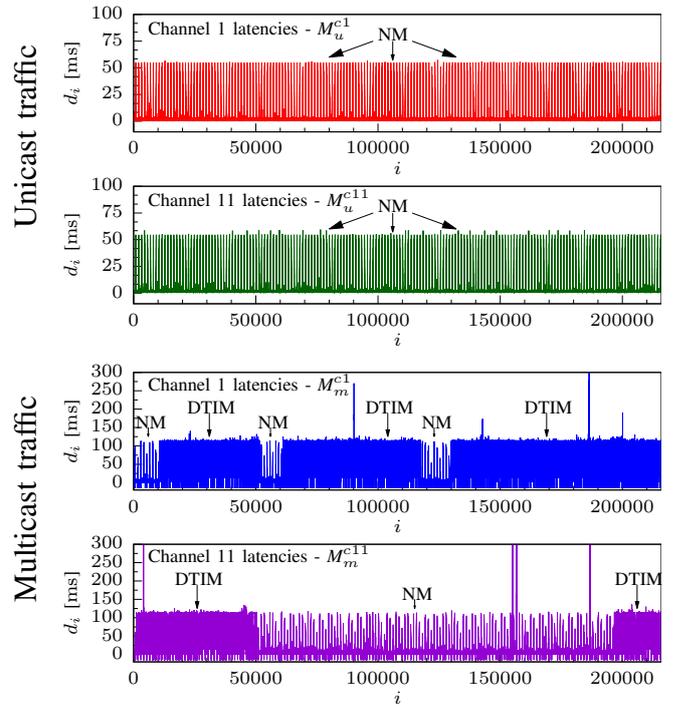

  \scriptsize
  \centering
  \include{FIG3-17-0869}
  \caption{End-to-end transmission latencies and packet losses in preliminary experiments for both unicast streams (top) and multicast streams (bottom).}
  \label{fig:baseline}
\end{figure}

\section{Configuration and preliminary optimization}
\label{sec:exp_software}
A first set of experiments were carried out by using the testbed ``out-of-the-box'', i.e., without any specific optimization.
As a matter of fact, conventional software modules and default equipment configuration resulted unsuitable for direct use in industrial applications. 
In particular, experiments highlighted the presence of interference affecting both channels of the redundant link, thus reducing expected performance.

\subsection{Preliminary Evaluation}
\label{sec:baseline}
Two experiments, each one lasting  one week, were performed to analyze the behavior of Wi-Fi seamless redundancy in a default configuration, using unicast and multicast services, respectively.
They were considered as the baseline.
Transmissions on the redundant link were modeled as a pair of concurrent synchronous streams, sent on channels $1$ and $11$.
The generation period $T_c$ was set to $\unit[100]{ms}$.
Fig.~\ref{fig:baseline} summarizes experimental data in two significant intervals.
The two topmost  timing diagrams ($\msg^{c1}_u$, $\msg^{c11}_u$) concern the unicast case, whereas those at the bottom ($\msg^{c1}_m$, $\msg^{c11}_m$) refer to multicast.
Each diagram spans over $6$ hours of uninterrupted  transmissions and shows the end-to-end latency $d_i$ experienced by packet $i$ on the related channel. 
In order to provide as much details as possible about transmissions, lost packets are visualized in the diagrams as negative latency values.

Two distinct interfering phenomena can be observed, which worsen communication determinism by causing packet latencies and losses.
The first one is due to the IEEE 802.11 Delivery Traffic Indication Map (DTIM) mechanism, which concerns energy saving. 
Its observable effects, corresponding to zones in diagrams for multicast streams $\msg^{c1}_m$ and $\msg^{c11}_m$ where interference is more dense, appear to be randomly distributed, because DTIM is activated when a STA that can enter the power-save state associates with the AP. 
In our experiments, since no authentication was required for the Basic Service Sets (BSS), any such STA entering the BSS range (e.g., mobile phones) might accidentally set DTIM on.
Since this mechanism only affects multicast traffic, experimental results for unicast streams $\msg^{c1}_u$ and $\msg^{c11}_u$ did not exhibit this behavior.
DTIM is analyzed in detail in Section~\ref{sec:DTIM}.
The second phenomenon is periodic and noticeably less intense than DTIM.
However, it affected both multicast and unicast transmissions for the whole experiment.
Investigations allowed us to conclude that it is due to a software module, running in the originating RSTA, which carries out a number of non-trivial functions, such as managing association/reassociation of the STAs with the AP and migration between adjacent BSSs (roaming). 
More details on this aspect are provided in Section~\ref{sub:nm}.

\subsection{DTIM mechanism}
\label{sec:DTIM}
\begin{figure}[t]
  \centering
  \includegraphics[width=.9\columnwidth]{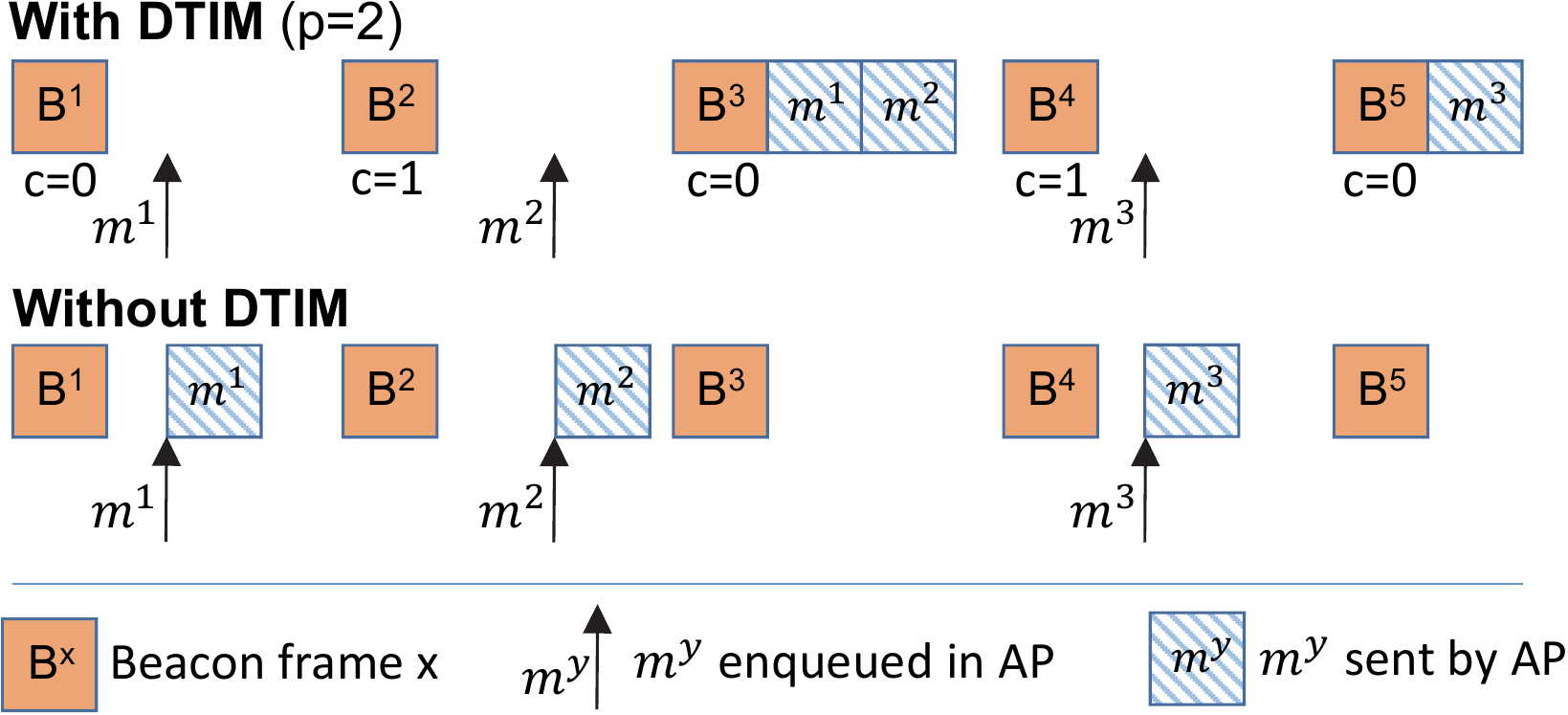}
  \caption{Example of multicast frame exchange (with and without DTIM).}
  \label{fig:DTIM_mechanism}
\end{figure}

According to the IEEE 802.11 standard, the DTIM mechanism is activated when at least one STA associated to the BSS enters the power-save mode. To ensure that all STAs are awake at the time the AP sends the multicast packets, they are buffered inside the AP and sent at specific time instants.
The AP notifies the awakening time to the STAs by means of two fields, \textit{DTIM period} ($p$) and \textit{DTIM count ($c$)}, included in the \textit{beacon} frame, which is cyclically sent by the AP every $T_\mathrm{beac}$ (the typical value is $\unit[102.4]{ms}$) and received by all STAs. Buffered multicast packets are sent out by the AP cyclically, with a period equal to $T_\mathrm{beac}*p$. 
The value $p$ is a configurable parameter of the AP (as $T_\mathrm{beac}$), and allows to set the sleeping period of the STAs.
The value $c$ is read by STAs to know how many beacons have still to elapse before the AP sends the buffered packets.
The value $c$ is decreased by $1$ by the AP on every beacon frame transmission and re-initialized to $c=p-1$ every time it reaches value $0$. In particular, $c=0$ means that multicast packets transmission will follow the current beacon. 
Operation of the DTIM mechanism is summarized in Fig.~\ref{fig:DTIM_mechanism}.

\subsubsection{Effects of DTIM on communication}
\begin{figure}[t]
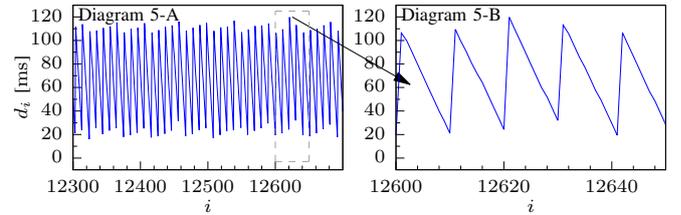

  \scriptsize
  \centering
  \include{FIG5-17-0869}
  \caption{Effects of the DTIM mechanism on end-to-end transmission latencies for multicast streams (unicast streams are not affected).}
  \label{fig:dtim}
\end{figure}

End-to-end latencies and losses of two multicast streams $\msg^{c1}_m$ and $\msg^{c11}_m$ have been logged during an additional experiment lasting one hour.
To provide higher resolution, packets were generated by the measurement task every $T_c=\unit[10]{ms}$.
Experimental data, concerning a time interval where DTIM was active, are shown in Fig.~\ref{fig:dtim}, with timing diagrams~\ref{fig:dtim}-A and \ref{fig:dtim}-B providing two different zoom levels. Latencies varied between $\unit[\sim\!15]{ms}$ and $\unit[\sim\!115]{ms}$, according to a saw-tooth law whose period is about $\unit[100]{ms}$.
This behavior is compatible with the DTIM mechanism: multicast packets, sent on the Ethernet port, reach the switch and then the AP, where they are buffered until the next beacon.
At the time of beacon transmission, a number of packets equal to about $T_\mathrm{beac}/T_c$ ($\sim\!10$ in our experiment) are queued in the AP.
They are sent on air back-to-back just after the beacon frame, and this results in close arrival times (typically, consecutive received packets are spaced by less than $\unit[1]{ms}$), whereas their transmission times are displaced by $T_c$.
This explains the shape of diagram~\ref{fig:dtim}-B and why, between adjacent beacons, latency decreases by $\sim\!\unit[9]{ms}$ per sample. 

DTIM may significantly worsen determinism of multicast packet streams.
When it is active, their measured transmission latencies suffered from a jitter that, when $p = 1$, is as high as one hundred $\unit[]{ms}$, and grows linearly for higher $p$ values.
This is hardly acceptable in real-time industrial applications.
Even worse, the probability that both channels of a redundant link are affected by DTIM at the same time is not negligible (see, e.g., bottom diagrams in Fig.~\ref{fig:baseline}).
As a consequence, effectiveness of seamless redundancy may be severely impaired.

\subsubsection{Possible remedies to DTIM issues}
A first, trivial rule to improve communication quality on the redundant link is to  improve the performance of each  physical channel separately.
In particular, a simple remedy to circumvent the DTIM problem is to hide the Service Set Identifier (SSID), a string embedded in beacon frames that identifies the BSS.
In this way, STAs not belonging to the control system are prevented from unintentionally associating with the AP. 
Unfortunately, this method only provides loose guarantees: in fact, a number of security attacks are known which target Wi-Fi networks configured this way.
A better and safer approach relies on the use of authentication.
Currently, the \emph{de facto} standard is IEEE 802.11i, which standardizes the security protocol popularly known as Wi-Fi Protected Access II (WPA2). 
By using WPA2, STAs not involved in the control network cannot authenticate with the AP, hence preventing DTIM interference.
Of course, only authentication is mandatory to avoid DTIM effects. 
However, since security attacks on industrial traffic may lead to harmful consequences, encryption is highly recommended.

In \cite{2016-ETFA-Guidelines}, experiments were carried out for measuring to what extent latencies $d_i$ are practically affected by WPA2 when using recent commercial adapters.
Results show that minimum and mean values ($d_{\operatorname{min}}$ and $\bar d$) were only marginally larger than an open system, the difference being in the order of a few $\unit[]{\mu s}$.
Worsening was slightly higher in terms of percentiles, but still negligible if compared to latency values.
Hence, it can be safely assumed that the overhead introduced for encryption is mostly irrelevant in the light of the benefits, which justifies the use of WPA2 to prevent unwanted DTIM effects.

\section{Interference inside devices}
Even when channels are set to operate on distinct frequencies (hence making frame transmissions on air totally independent, in theory), other sources of interference may affect communications in a joint fashion.
This is because redundant transmission paths are not completely disjoint, but include overlapping portions located inside devices.
In general, interference in the communication hardware (physical and MAC layers) is negligible, as wireless adapters are replicated and operate independently.
However, a small amount of residual interference still exists, because operations involved in packet transmission and reception are (in part) managed in software by the same hardware components, that is, CPU, memory (RAM), and system bus (e.g., PCIe).
\begin{figure*}[t]
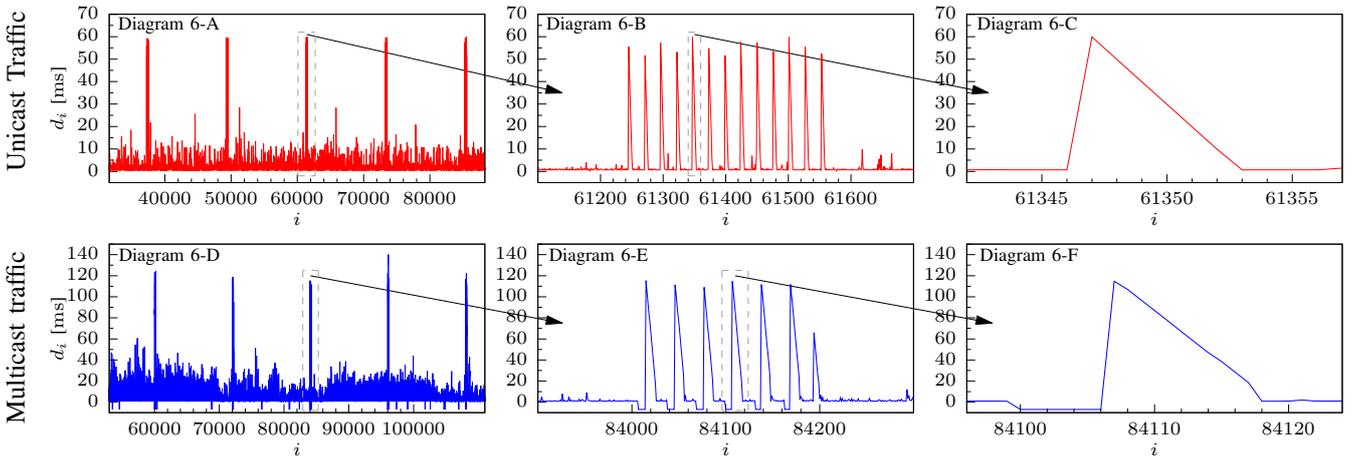

  \scriptsize
  \centering
  \include{FIG6-17-0869}
  \caption{Effects of the network manager service on end-to-end transmission latencies and losses for both unicast streams (top) and multicast streams (bottom).}
  \label{fig:nm}
\end{figure*}

Exploiting concurrent tasks on multi-core CPUs may lessen this problem.
Moreover, when fast CPUs are employed, as in modern PCs, this interference only causes short additional delays and jitters (in the order of few tens $\unit[]{\mu s}$).
For this reason, these contributions have not been considered explicitly in the remaining part of this paper.
The same is no longer true when APs are taken into account, as they are usually based on embedded microcontrollers characterized by lower performance.
Hence, internal interference in APs can affect the quality of communication much more than in the PC.

Besides the hardware components shared between redundant paths, software tasks may also cause non-negligible joint interference, when they repeatedly act on wireless adapters at the same time, e.g., to carry out management activities.
A significant example of such a kind of behavior is described in the following section.

\subsection{Internal interference due to network management services}
\label{sub:nm}
To achieve independence between channels in a redundant link, wireless adapters should never be engaged in concurrent activities triggered by tasks other than the originator and recipient of packets.
Unfortunately, this condition may occasionally occur in real systems.
For instance, Linux includes a \emph{network manager} (NM) service, which periodically scans all wireless channels to discover nearby BSSs (from a practical viewpoint, it looks for APs the STA may connect to).
Similar services are foreseen in most other operating systems as well. 
The NM service is essential to ensure an adequate level of reliability.
In fact, it allows STAs to reconnect to the AP when, because of disturbance conditions lasting long enough, they get disassociated.
Moreover, NM also supports mobility between BSSs, hence achieving larger coverage areas. 
The ability to automatically reconnect STAs is practically mandatory in automation systems, to avoid situations where communications are permanently broken and human intervention is required.
On the contrary, mobility is needed only in specific industrial scenarios, when movements of STAs may exceed the coverage area of a single BSS, e.g., in the case of automated guided vehicles (AGV) and drones.

Periodic channel scanning, carried out by NM for all Wi-Fi adapters installed in the device, causes non-negligible performance degradation, as experimental results in Fig.~\ref{fig:baseline} clearly highlight.
In fact, while this task is in progress, application data cannot be sent and not even received.
This implies that packets to be transmitted are temporarily buffered, and sent on air only when scanning is concluded.
Unavoidably, this introduces delays for each single adapter.

In systems exploiting seamless redundancy, this issue can be prevented by purposely displacing in time the NM operations on different channels.
Unfortunately, our experiments showed that, in conventional platforms like Linux PCs, scanning on different adapters took place at the same time.
This makes interference on channels no longer independent, and worsens the communication quality of the redundant link consequently.

To evaluate NM effects, two experiments were performed, where the DTIM issue was prevented, by configuring the adapters to operate on channels $1$ and $11$, and using both unicast ($\msg^{c1}_u$, $\msg^{c11}_u$) and multicast ($\msg^{c1}_m$, $\msg^{c11}_m$) streams.
Each experiment lasted one hour and, in both cases, $T_c=\unit[10]{ms}$.
Fig.~\ref{fig:nm} shows end-to-end latencies and losses for the adapter tuned on channel $1$.
Timing diagrams in the upper part of the figure concern unicast frames, while those in the lower part are about multicast.
Results for the second adapter are similar, and hence they are not shown in the figure.

\subsubsection{Unicast transmission}
Diagram~\ref{fig:nm}-A (top-left) refers to a $10$-minute-wide interval. 
Periodic interference, in the form of peaks of transmission latency $d_i$, appeared every two minutes for the whole duration of the experiment. 
They are due to the NM service running in the PC.
\mbox{Diagram~\ref{fig:nm}-B} (top-middle) shows a single scan performed by NM across all channels.
Its duration, measured between the outermost peaks, is $\unit[\sim\!3]{s}$ ($308$ packets are affected).
It includes $13$ peaks, corresponding to the probe operations NM carried out on each one of the $13$ channels legally allowed for \mbox{Wi-Fi} in Europe, according to the European Telecommunications Standards Institute (ETSI) regulations \cite{2015-std-etsien300328}.
Finally, diagram~\ref{fig:nm}-C (top-right) shows an expanded view of a single peak, 
and depicts the effects of a specific probing operation.

When NM tunes the RF module of the adapter used to transmit stream $\msg^{c1}_u$ on a new channel (e.g., $5$), the measurement task temporarily loses its ability to send the related packets, which are temporarily buffered in the STA's adapter.
If the number of packets generated during NM activity exceeds the transmission buffer capacity, some packets may be possibly discarded.
As soon as NM switches the RF module back to channel $1$ and re-enables communication, buffered packets are sent on air, in the same order as they were queued.
Probing in diagram~\ref{fig:nm}-C affected $6$ consecutive packets: the ``older'' packet suffers from the largest latency ($\unit[\sim\!60]{ms}$), and delays decrease on each following packet by about $T_c$ ($\unit[10]{ms}$) minus the time to transmit one frame on air, which in our setup was in the order of one hundred $\unit[]{\mu s}$.
Not all peaks in diagram~\ref{fig:nm}-B have exactly the same height, because the generation process of $\msg^{c1}_u$ and the activation interval of NM are not synchronized.

\subsubsection{Multicast transmission}
Results for multicast packets are similar to the unicast case, with some peculiar differences due to the fact that, in our setup, they flow from APs to the STAs.
When NM is active, packets are dropped because the RF module in the STA is listening to a different channel.
Unlike the unicast case, where this issue is counteracted by the AP using retransmissions, no retry is carried out for multicast frames as they are unconfirmed.
As for the baseline case, in diagrams~\ref{fig:nm}-D, \ref{fig:nm}-E, and \ref{fig:nm}-F, dropped packets are represented as negative latency values.

Diagram~\ref{fig:nm}-E concerns a complete scan of all channels.
Only $7$ peaks are observed instead of $13$: this is because, before switching the RF module frequency, NM temporarily activates the DTIM mechanism in the AP, in an attempt not to lose received multicast packets.
This causes packets, generated by the measurement task in a time interval comprising scans on several channels (two, in the case of diagram~\ref{fig:nm}-F), to be buffered by the AP and sent immediately after a beacon.
For this reason the height of peaks in diagrams referred to multicast streams is about twice the unicast case.
The last peak is shorter ($\unit[\sim 60]{ms}$) because only one channel is concerned.
Although the particular shape of diagrams depends on the specific implementation of NM in Linux, similar amounts of interference are nevertheless expected in different operating systems to carry out the related functionality.

\subsubsection{Possible remedies to NM interference}
Although a network scanning service has to be mandatorily implemented in IEEE 802.11, scanning sequences are not typically triggered automatically by the adapter, but are invoked on demand by the NM.
Disabling NM \cite{2014-ETFA-trama} is the easiest solution to avoid above described issues.
This has been done in all experiments reported in the remaining part of this paper, so that results are not affected by NM interference.
However, the NM service is fundamental for industrial Wi-Fi networks, where a permanent loss of connectivity cannot be tolerated.
In fact, when a STA experiences a disconnection from the AP, it is unable to reassociate without the NM intervention.
Both sides of a wireless link (STA and AP) may force a disconnection, if no communication takes place for a certain lapse of time.
This condition occurs frequently for mobile nodes (i.e., those allowed to migrate to areas covered by other APs), but it may also happen for unmovable nodes (including those whose movements are confined inside the coverage area of a single AP), due, e.g., to prolonged interference or disturbance on air.
Solutions other than disabling NM are thus needed.

If properly exploited, network redundancy can offer tangible advantages for NM operations.
In particular, it can lessen the impact this service has on data transfers, as well as the likelihood that all channels become disconnected at the same time.
To this aim, specific NM implementations, tailored to such application contexts, have to be developed, which take into account their specific characteristics and, in particular, whether devices are unmovable or mobile.
Channel scanning is not strictly required for unmovable STAs, in which case network parameters can be configured statically.
As an alternative, the discovery procedure can be executed only once, at network startup.
In both cases, NM has only to monitor the STA connection state, and reassociate to the same AP in the shortest possible time when a disconnection occurs.

When mobility is required, NM has to periodically scan all channels involved in the redundant link to acquire information about nearby APs. 
Importantly, this operation must never be carried out on the two adapters at the same time.
In a similar way, when a device wishes/needs to migrate to another AP, NM has to carry out reassociation on one channel at a time. 
Only when this operation is concluded on one adapter, it can be started on the other adapter too.
This ensures that, during migration, communication is never interrupted.
A number of heuristics can be devised to manage roaming, aimed at deciding timings and the channel order used for migration, based on information provided by the MAC, e.g., the signal-to-noise ratio or the PLR on each channel. 
These peculiar aspects fall outside the scope of this paper and are left for future investigations.

\subsection{Internal interference in network equipment}
\label{sec:NetworkComponents}

As shown in Fig.~\ref{fig:testbed}, the simplest way to set up an infrastructure RBSS is deploying a pair of APs (configuration $\APtwo$).
As an alternative, a single simultaneous dual-band AP (an AP provided with two radio blocks and two MAC entities, operating on distinct channels at the same time) can be used (configuration $\APone$).
This permits the roles of the two APs in Fig.~\ref{fig:testbed} to be functionally merged in the same device.
Unlike RSTAs implemented on PCs, where the code of the protocol stack is often available, and can be customized to cope with phenomena causing joint interference on redundant paths, AP firmware is seldom accessible, especially for COTS devices.

To evaluate the effects of internal interference in the network equipment, we configured a single AP (NETGEAR WAC120) to operate simultaneously on channel $1$ in the $\unit[2.4]{GHz}$ band and $165$ in the $\unit[5]{GHz}$ band.
A cyclic unicast traffic ($\msg^{c1}_u$, $\msg^{c165}_u$) with period $T_c=\unit[100]{ms}$ and directed to the AP was generated by the PC.
The experiment lasted $6$ hours and results are reported in the top rows of Table~\ref{tab:diversity}, labeled $\APone$ (clearly, end-to-end latencies were considered).
The very same experiment was repeated using one AP per channel (two identical NETGEAR WAC120 were employed in this case). Results are included in the second set of rows, labeled $\APtwo$.

\begin{figure}[t]
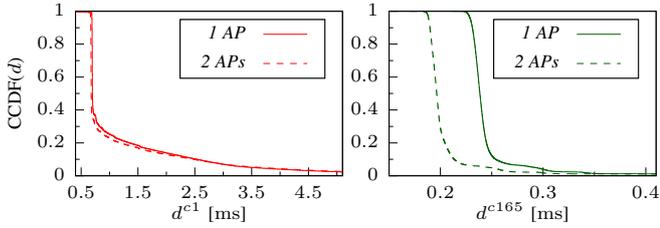

  \scriptsize
  \centering
  \include{FIG7-17-0869}
  \caption{CCDF of end-to-end latencies on channels $1$ ($d^{c1}$) and $165$ ($d^{c165}$).}
  \label{fig:diversityccdf}
\end{figure}

\begin{table}[t]
  \caption{Effects of network infrastructure on seamless redundancy: one simultaneous dual-band AP vs. two single-band APs.}
  \label{tab:diversity}
  \scriptsize
  \begin{center}
    \renewcommand{\arraystretch}{1.4}
    \tabcolsep=0.20cm
    \begin{tabular}{cc|ccccc|c}
      Exp. & Stream & $\bar{d}$ & $\sigma_{d}$ & $d_{p99.9}$ & $d_{p99.99}$  & $d_{\operatorname{max}}$ & $\mathrm{PLR}$\\
      & & \multicolumn{5}{c|}{[$\unit[]{ms}$]} & [$\unit[]{\%}$] \\
      \hline \hline
      \multirow{3}{*}{\begin{sideways}$\APone$\end{sideways}}
      & $\msg^{c1}_u$          & 1.267 & 1.685 & 20.064 & 27.775 & 41.044 & 0.0005 \\
      & $\msg^{c165}_u$         & 0.346 & 1.179 & 18.639 & 20.425 & 21.466 & 0.0 \\
      & $\msg^{c1 \cup c165}_u$  & 0.346 & 1.179 & \textbf{18.639} & \textbf{20.425} & 21.466 & 0.0 \\
      \hline
      \multirow{3}{*}{\begin{sideways}$\APtwo$\end{sideways}}
      & $\msg^{c1}_u$          & 1.236 & 1.730 & 20.071 & 27.483 & 40.939 & 0.0005 \\
      & $\msg^{c165}_u$         & 0.304 & 1.173 & 18.457 & 20.241 & 20.430 & 0.0 \\
      & $\msg^{c1 \cup c165}_u$  & 0.210 & 0.133 &  \textbf{2.077} &  \textbf{5.110} & 10.839 & 0.0 \\
      \hline
    \end{tabular}
  \end{center}
\end{table}

\begin{figure}[t]
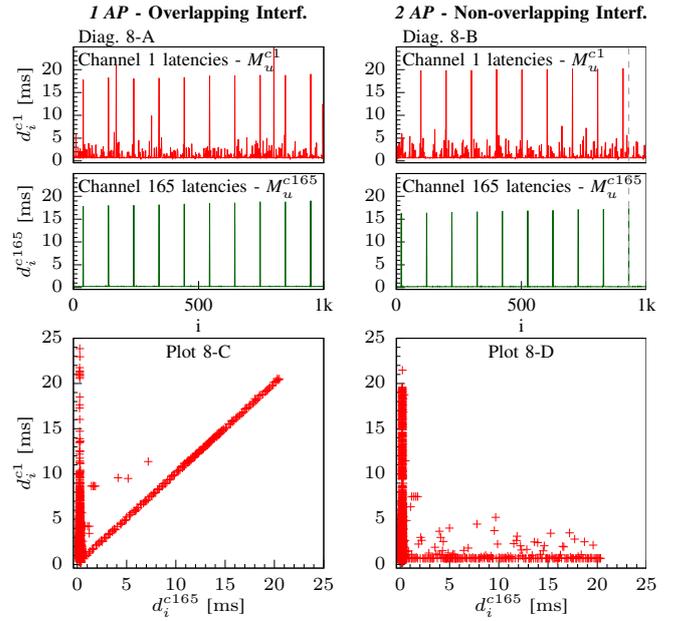

  \scriptsize
  \centering
  \include{FIG8-17-0869}
  \caption{End-to-end transmission latency on channels 1 ($d_i^{c1}$) and 165 ($d_i^{c165}$), shown using timing diagrams (top) and scatter plots (bottom), for $\APone$ and $\APtwo$ configurations.}
  \label{fig:diversity}
\end{figure}

\subsubsection{Forwarding delays}
$\msg^{c1}_u$ statistical indexes  are similar in the two experiments, and the same holds for $\msg^{c165}_u$.
Differences were due in part to the fact that the wireless medium is not stationary (i.e., channel characteristics change over time).
Main causes are disturbance (caused by electromagnetic noise, moving objects, multipath fading, attenuation, etc.) and interfering traffic generated by nearby Wi-Fi networks, both of which are out of the experimenter's control.
For the $\APone$ configuration, only $0.18\%$ of the packets arrived first on channel $1$, which causes statistical indexes of stream $\msg^{c1 \cup c165}_u$ (which  models the redundant link) to be almost the same as for channel $165$ (see second and third lines of Table~\ref{tab:diversity}).

Fig.~\ref{fig:diversityccdf} shows the \emph{complementary cumulative distribution functions} (CCDF), measured in experiments $\APone$ and $\APtwo$, for both channels $1$ and $165$.
Overall, latencies in the $\unit[2.4]{GHz}$ band are worse than in the $\unit[5]{GHz}$ band, mostly because of the higher interference on air.
More in detail, the CCDFs of ${d}^{c1}$ (plot on the left) are roughly the same for the two experiments.
Instead, the CCDF of ${d}^{c165}$ (plot on the right) in $\APone$ is shifted to the right by $\unit[\sim\!40]{\mu s}$ with respect to $\APtwo$, likely because of the additional overhead for managing operations in both bands. 
Thus, when a single AP is employed, the time taken to forward packets received on the $\unit[5]{GHz}$ band  increases, on average, by the same amount, which directly worsens seamless redundancy performance.
This is also confirmed by mean latency values on channel $165$ ($\bar{d}^{c165}$), equal to $\unit[0.346]{\mu s}$ and $\unit[0.304]{\mu s}$ for configurations $\APone$  and $\APtwo$, respectively.

\subsubsection{Joint interference}
A more serious phenomenon that affects seamless redundancy is shown in the timing diagrams in the upper part of Fig.~\ref{fig:diversity}.
As can be seen, some internal activity is cyclically executed in our APs every $\unit[10]{s}$, which interferes with packet forwarding.
Packets reaching a busy AP suffer from additional delays that, in the worst case, can be as long as $\unit[20]{ms}$.
If the same AP manages both wireless networks ($\APone$), this interference always occurs at the same time on both channels (see diagrams~\ref{fig:diversity}-A), hence affecting the redundant link to the same extent.
Conversely, interference seldom occurs in a joint manner when two distinct APs are employed ($\APtwo$), since their operation is not coordinated (see diagrams~\ref{fig:diversity}-B, where peaks are displaced in time).

Statistical dependence can be better analyzed by means of the scatter plots in the bottom part of Fig.~\ref{fig:diversity}, where each point represents a packet and its $XY$ coordinates correspond to latencies $d_i^{c165}$ and $d_i^{c1}$ experienced on channels $165$ and $1$, respectively.
Most packets suffer from negligible delays, and the related points fall close to the plot origin.
As plot~\ref{fig:diversity}-D shows, in the $\APtwo$ configuration several points are placed near the $X$ and $Y$ axes.
This is because $d_i^{c165}$ and $d_i^{c1}$ are independent and, as shown in diagrams~\ref{fig:diversity}-B, time intervals where interference occurs seldom overlap, since the time bases in the two APs are not synchronized.
Points lying near the identity line ($X=Y$) in plot~\ref{fig:diversity}-C, which concerns the $\APone$ configuration, correspond to packets that experienced similar, non-negligible delays on both channels (i.e., when $d_i^{c165} \simeq d_i^{c1}$).
They confirm the presence of joint interference, which delays the copies of the same packet by the same amount of time on both channels.
When this behavior is prevented, as in the $\APtwo$ case, statistical indexes about the latency on the redundant link improve noticeably. 
For example, as shown in Table~\ref{tab:diversity}, the $99.9$ and $99.99$ percentiles decrease from $18.639$ and $\unit[20.425]{ms}$ in the $\APone$ configuration to  $2.077$ and $\unit[5.110]{ms}$ for the $\APtwo$ configuration, respectively.

\subsubsection{Possible remedies to AP internal interference}
The above examples put into evidence some aspects, due to infrastructure network equipment, which might impair the effectiveness of seamless redundancy.
An obvious suggestion is that, in (soft) real-time applications, devices have to be properly characterized before their actual deployment.
Unfortunately, this can result in a complex task and takes significant time and effort.
A slightly more expensive alternative is relying on distinct APs to reduce the common parts of redundant paths, so that the likelihood that interfering phenomena affect all copies of the same packet decreases.

\section{Interference on air}
\label{sec:DeterminingThePresenceOfACI}
Experiments with real devices revealed that another kind of interference exists for physical channels, which could affect frame transmissions on air.
An in-depth analysis highlighted that a residual statistical dependence was caused by the relative positions of antennas which, when placed close to one another, induced a phenomenon known as Adjacent Channels Interference (ACI) \cite{2008-IWCMCC-ACI}.
In our setup, wireless adapters were fastened to two PCIe slots of the same PC, so that antennas were located a couple of centimeters apart.
This condition is not so unrealistic, since real industrial devices are typically provided with small-sized enclosures, and antennas can not always be separated by means of cables.

\subsection{Evaluation of Adjacent Channel Interference}
To evaluate ACI, experiments based on unicast traffic were carried out using the setup with two APs of Fig.~\ref{fig:testbed}, after NM interference in the PC was prevented.
Link latencies $d^{\prime}$ measured on ACK arrivals were considered.
The packet loss ratio was determined both on the PC's Ethernet port ($\mathrm{PLR}$) and by counting packets for which no confirmation (ACK frame) was received by \mbox{Wi-Fi} adapters ($\mathrm{PLR}^\prime$).
Since ACK is not returned for lost frames, $\mathrm{PLR}^\prime \geq \mathrm{PLR}$.
However, in typical operating conditions, the likelihood that, for a given packet transmission, all the related ACK frames are corrupted is very low, and so $\mathrm{PLR}^\prime \simeq \mathrm{PLR}$.
To make results clearer, we disabled MAC layer retransmissions (i.e., we relied on one-shot acknowledged frames) and selected a fixed $\unit[54]{Mbit/s}$ bit rate.

In the following, the two physical channels are conventionally referred to as \textit{channel-under-test} ($\mathrm{M}$), on which measurements about packets were collected, and \textit{interfering channel} ($\mathrm{I}$), while the related adapters are denoted STA$_\mathrm{M}$ and STA$_\mathrm{I}$, respectively.
STA$_\mathrm{M}$ was set to channel $165$ in the $\unit[5]{GHz}$ band, while the operating frequency of STA$_\mathrm{I}$ was varied.
Packets were repeatedly transmitted on $\mathrm{M}$ by the measurement task with period $T_c = \unit[100]{ms}$.
On every second packet on $\mathrm{M}$, a similar packet was also sent on $\mathrm{I}$.
The sequence of packets on $\mathrm{M}$, defined as $\{\mes^{\overline{\ACI}}_1, \mes^{\ACI}_2, \mes^{\overline{\ACI}}_3, \mes^{\ACI}_4, ...\}$, consists of two distinct subsets, namely, $\overline{\ACI}=\{\mes^{\overline{\ACI}}_1, \mes^{\overline{\ACI}}_3,...\}$, which contains packets sent only on the channel-under-test, and  $\ACI=\{\mes^{\ACI}_2, \mes^{\ACI}_4,...\}$, which includes those packets concurrently sent on both channels.
Interleaving simplex and duplex streams allows to reliably compare ACI effects in almost the same environmental conditions, despite the wireless spectrum varies unpredictably over time.
Statistics on experimental samples in the two sets were compared to determine the impact of ACI on latencies and losses.

\subsection{ACI Effects on Seamless Redundancy}
Not all ACI effects on packet transmission can be easily predicted, because some of them depend on very specific behaviors of the physical layer. 
Generally speaking, interfering frames sent on non-overlapping adjacent channels by nearby antennas 
may interact with the Clear Channel Assessment (CCA) mechanism in such a way that the originator could wrongly sense the channel as busy, even when it is idle.
More rarely, they might even corrupt ongoing frame receptions.
Two relevant (although not exhaustive) examples of cross-channel interaction, we termed \textit{frame delaying} and \textit{ACK frame collision}, have been experimentally analyzed.

\begin{figure}[t]
  \scriptsize
  \centering
  \includegraphics[width=0.8\columnwidth]{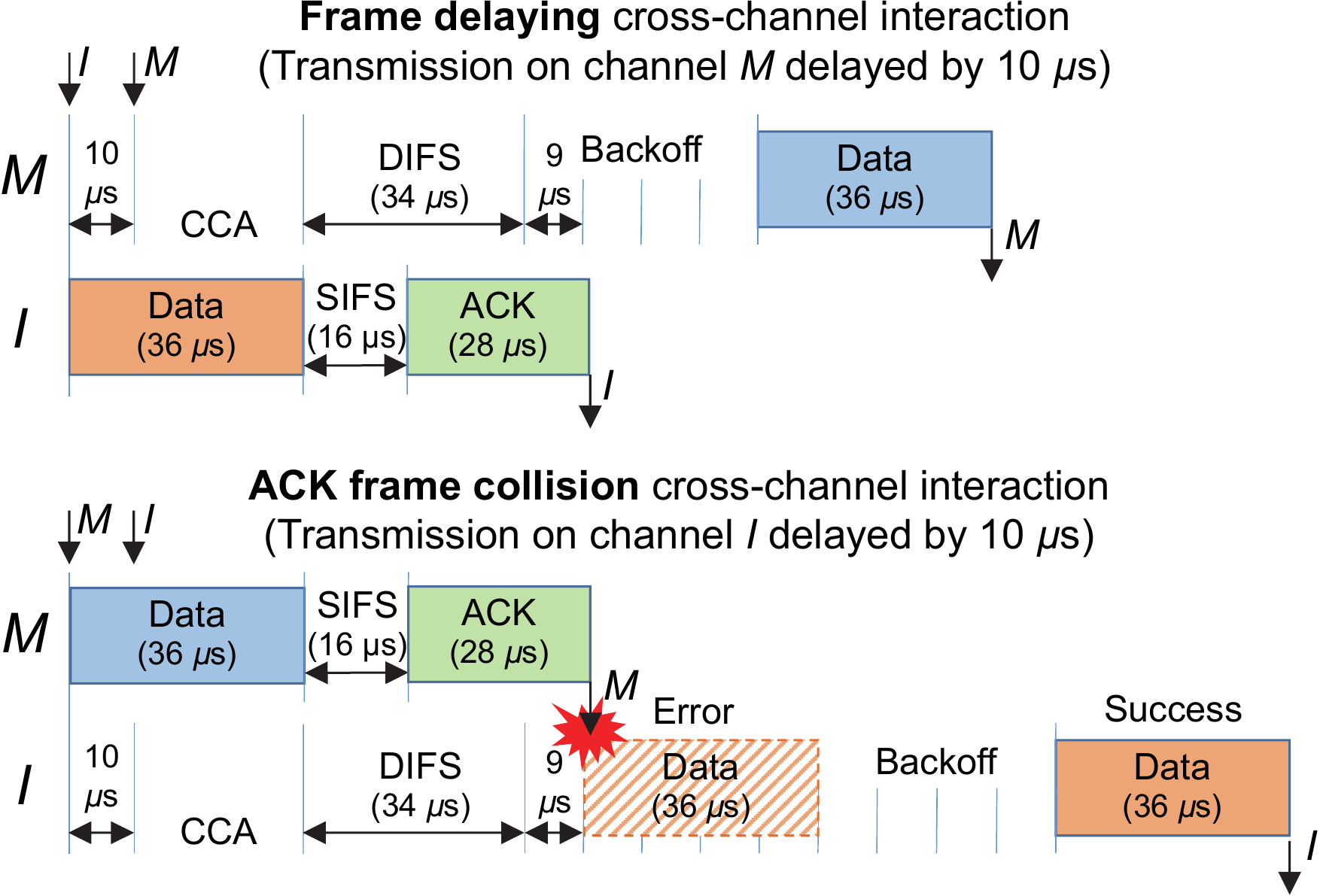}
  \caption{ACI effects on transmission: frame delaying and ACK frame collision.}
  \label{fig:aci}
\end{figure}

\subsubsection{Frame delaying}
\begin{table*}[t]
  \caption{Adjacent channel interference by varying the interfering channel: frame delaying and ACK frame collision effects.}
  \label{tab:ACI_results_new}
  \footnotesize
  \begin{center}
    \tabcolsep=0.11cm
    \begin{tabular}{cc||lllllll|ll||lllllll|ll}
      \multicolumn{20}{c}{\textbf{Frame delaying} cross-channel interaction (transmissions on channel $\mathrm{M}$ delayed by $\unit[10]{\mu s}$)} \\
                  &        & \multicolumn{9}{c||}{Samples suffering from ACI (channel 165, $\ACI$ set)} & \multicolumn{9}{c}{Samples unaffected by ACI (channel 165, $\overline{\ACI}$ set)} \\
      Ch.       & Ch.   & $\bar{d^\prime}$ & $\sigma_{d^\prime}$ & $d^\prime_{\operatorname{min}}$ & $d^\prime_{p50}$ & $d^\prime_{p99}$ & $d^\prime_{p99.9}$  & $d^\prime_{\operatorname{max}}$ & $\mathrm{PLR}^\prime$ & $\mathrm{PLR}$ & $\bar{d^\prime}$ & $\sigma_{d^\prime}$ & $d^\prime_{\operatorname{min}}$ & $d^\prime_{p50}$ & $d^\prime_{p99}$ & $d^\prime_{p99.9}$  & $d^\prime_{\operatorname{max}}$ & $\mathrm{PLR}^\prime$ & $\mathrm{PLR}$ \\
      $\mathrm{M}$ & $\mathrm{I}$ & \multicolumn{7}{c|}{[$\unit[]{ms}$]} & \multicolumn{2}{c||}{[\%]} & \multicolumn{7}{c|}{[$\unit[]{ms}$]} & \multicolumn{2}{c}{[\%]} \\
      \hline
      \hline
      165 & 161 &    \textbf{0.265} &     0.645 &     0.059 &     \textbf{0.176} &     \textbf{5.369} &     5.526 &     5.728 &     0.089 & 0.089
                &    \textbf{0.070} &     0.059 &     0.066 &     \textbf{0.068} &     \textbf{0.080} &     0.289 &     4.376 &     0.133 & 0.133 \textbf{(a)}\\
      165 & 157 &    \textbf{0.246} &     0.536 &     0.060 &     \textbf{0.177} &     \textbf{3.702} &     5.630 &     5.913 &     0.056 & 0.056
                &    \textbf{0.070} &     0.049 &     0.061 &     \textbf{0.068} &     \textbf{0.074} &     0.285 &     3.926 &     0.278 & 0.267 \\
      165 & 153 &    \textbf{0.088} &     0.128 &     0.057 &     \textbf{0.070} &     \textbf{0.240} &     2.662 &     4.801 &     0.050 & 0.050
                &    \textbf{0.069} &     0.014 &     0.061 &     \textbf{0.068} &     \textbf{0.077} &     0.263 &     1.104 &     0.011 & 0.011 \\
      165 & 149 &    \textbf{0.113} &     0.349 &     0.058 &     \textbf{0.069} &     \textbf{2.578} &     5.017 &     5.311 &     0.028 & 0.017
                &    \textbf{0.069} &     0.030 &     0.061 &     \textbf{0.068} &     \textbf{0.074} &     0.255 &     3.627 &     0.017 & 0.017 \\
      165 & 48  &    \textbf{0.073} &     0.057 &     0.059 &     \textbf{0.070} &     \textbf{0.085} &     0.274 &     5.456 &     0.011 & 0.011
                &    \textbf{0.069} &     0.018 &     0.066 &     \textbf{0.068} &     \textbf{0.077} &     0.301 &     1.403 &     0.011 & 0.006 \\
      165 & 36  &    \textbf{0.073} &     0.081 &     0.059 &     \textbf{0.069} &     \textbf{0.085} &     0.319 &     5.248 &     0.022 & 0.022
                &    \textbf{0.069} &     0.015 &     0.062 &     \textbf{0.068} &     \textbf{0.073} &     0.272 &     1.232 &     0.039 & 0.039 \\
      165 & 1   &    \textbf{0.070} &     0.013 &     0.066 &     \textbf{0.069} &     \textbf{0.085} &     0.262 &     1.003 &     0.039 & 0.039
                &    \textbf{0.070} &     0.013 &     0.066 &     \textbf{0.069} &     \textbf{0.085} &     0.229 &     1.066 &     0.122 & 0.122 \\
      \hline
      \\ \\ \\
      \multicolumn{20}{c}{\textbf{ACK frame collision} cross-channel interaction (transmissions on channel $\mathrm{I}$ delayed by $\unit[10]{\mu s}$)} \\
                  &        & \multicolumn{9}{c||}{Samples suffering from ACI (channel 165, $\ACI$ set)} & \multicolumn{9}{c}{Samples unaffected by ACI (channel 165, $\overline{\ACI}$ set)} \\
      Ch.       & Ch.   & $\bar{d^\prime}$ & $\sigma_{d^\prime}$ & $d^\prime_{\operatorname{min}}$ & $d^\prime_{p50}$ & $d^\prime_{p99}$ & $d^\prime_{p99.9}$  & $d^\prime_{\operatorname{max}}$ & $\mathrm{PLR}^\prime$ & $\mathrm{PLR}$ & $\bar{d^\prime}$ & $\sigma_{d^\prime}$ & $d^\prime_{\operatorname{min}}$ & $d^\prime_{p50}$ & $d^\prime_{p99}$ & $d^\prime_{p99.9}$  & $d^\prime_{\operatorname{max}}$ & $\mathrm{PLR}^\prime$ & $\mathrm{PLR}$ \\
      $\mathrm{M}$ & $\mathrm{I}$ & \multicolumn{7}{c|}{[$\unit[]{ms}$]} & \multicolumn{2}{c||}{[\%]} & \multicolumn{7}{c|}{[$\unit[]{ms}$]} & \multicolumn{2}{c}{[\%]} \\
      \hline
      \hline
      165 & 161 &    0.069 &     0.033 &     0.060 &     0.068 &     0.075 &     0.355 &     4.902 &     \textbf{0.334} & \textbf{0.012}
                &    0.069 &     0.032 &     0.065 &     0.068 &     0.074 &     0.285 &     4.889 &     \textbf{0.019} & 0.014 \\
      165 &  36 &    0.070 &     0.011 &     0.058 &     0.069 &     0.077 &     0.251 &     1.484 &     \textbf{0.192} & \textbf{0.014}
                &    0.069 &     0.011 &     0.057 &     0.068 &     0.075 &     0.249 &     1.685 &     \textbf{0.014} & 0.014 \\
      165 &   1 &    0.069 &     0.135 &     0.058 &     0.068 &     0.075 &     0.244 &    44.425 &     \textbf{0.007} & \textbf{0.007}
                &    0.069 &     0.014 &     0.057 &     0.068 &     0.075 &     0.246 &     2.994 &     \textbf{0.016} & 0.016 \\
      \hline
    \end{tabular}
  \end{center}
\end{table*}

To make this phenomenon more apparent, transmissions on $\mathrm{I}$ were set to occur slightly in advance ($\unit[10]{\mu s}$) with respect to the corresponding ones on $\mathrm{M}$.
This situation is sketched in the topmost diagram of Fig.~\ref{fig:aci}.
When a request is issued for a frame transmission on $\mathrm{M}$, STA$_\mathrm{M}$ founds the channel busy, due to the ongoing transmission on $\mathrm{I}$.
As soon as STA$_\mathrm{M}$ senses the channel idle, it waits for a DIFS and the following backoff before starting transmission on air.
As a consequence, frame transmission on $\mathrm{M}$ is postponed by the duration of the frame on $\mathrm{I}$ plus a DIFS and the random backoff selected by STA$_\mathrm{M}$.
Interestingly, channel $\mathrm{M}$ is sensed idle by STA$_\mathrm{M}$ upon completion of the interfering data frame, and not the related ACK, because the latter comes from the AP, which is located some meters apart from the PC.

The experiment consisted in a number of trials, each one lasting one hour, carried out by changing the interfering channel $\mathrm{I}$ ($161$, $157$, $153$, $149$, $48$, and $36$ in the $\unit[5]{GHz}$ band, and $1$ in the $\unit[2.4]{GHz}$ band).
This permitted to evaluate how much the channel-under-test $\mathrm{M}$ ($165$) is affected by ACI due to channel $\mathrm{I}$ vs. the offset between channel frequencies.
Results are reported in the upper part of Table~\ref{tab:ACI_results_new}, where each row concerns a trial.
Results clearly show that ACI effects appear, to varying degrees, for almost all channels in the $\unit[5]{GHz}$ band. 

\begin{figure}[]
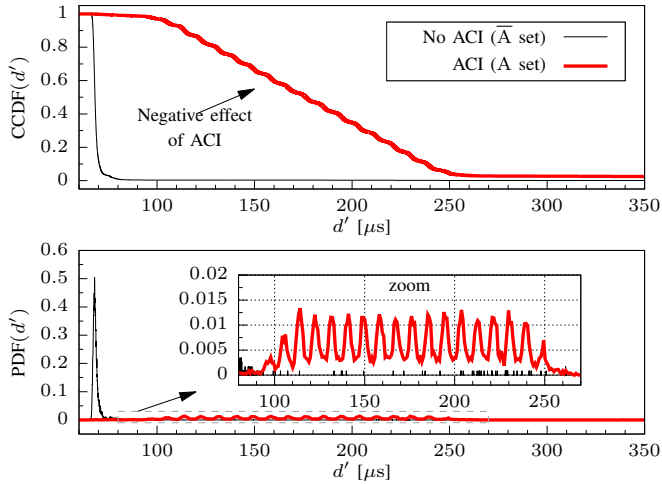

  \scriptsize
  \centering
  \include{FIG10-17-0869}
  \caption{CCDF and PDF of measured link latencies for experiment (a) in Table~\ref{tab:ACI_results_new} (channel $165$ vs. $161$) to highlight the frame delaying effect.}
  \label{fig:aci_dist_ccdf}
\end{figure}

The closer the channels are, the more the latency increases.
For instance, when $\mathrm{I}$ is set to channel $161$, the average link latency $\bar d^{\prime M}_\ACI$ for packets on $\mathrm{M}$ in the $\ACI$ set is $\unit[0.265]{ms}$, much larger than in the $\overline{\ACI}$ set, where $\bar d^{\prime M}_{\overline \ACI} = \unit[0.070]{ms}$.
Even bigger differences between the two sets can be evinced from percentiles.
The CCDF and the \textit{probability density function} (PDF) of link latencies $\bar d^{\prime M}_\ACI$ and $\bar d^{\prime M}_{\overline \ACI}$ for this specific experiment are reported in Fig.~\ref{fig:aci_dist_ccdf}.
They show two important facts: first, the CCDF changed sensibly because of ACI, with a tangible increase in latencies;
second, the PDF reveals that, as expected, the majority of the packets in the $\ACI$ set are delayed by the backoff interval, whose effects correspond to the peaks in the zoomed area \cite{2007-TII-PDF}.

ACI effects gradually decrease when the frequency offset between channels $\mathrm{M}$ and $\mathrm{I}$ increases, and disappear completely when $\mathrm{I}$ is set to operate on a different band ($\unit[2.4]{GHz}$).

\subsubsection{ACK frame collision}
\begin{table*}[t]
  \caption{Results after optimization (unicast transmissions).}
  \label{tab:final_results_unicast}
  \scriptsize
  \begin{center}
    \begin{tabular}{ccc||lllll|l||lllll|l}
          &        &        & \multicolumn{6}{c||}{Statistics derived from the PC Ethernet port} & \multicolumn{6}{c}{Statistics derived from ACK frame arrivals} \\
      \multicolumn{2}{c}{Experiment} & Stream & $\bar{d}$ & $\sigma_{d}$ & $d_{p99.9}$ & $d_{p99.99}$  & $d_{\operatorname{max}}$ & $\mathrm{PLR}$ & $\bar{d^\prime}$ & $\sigma_{d^\prime}$ & $d^\prime_{p99.9}$ & $d^\prime_{p99.99}$  & $d^\prime_{\operatorname{max}}$ & $\mathrm{PLR}^\prime$ \\
      &  &  & \multicolumn{5}{c|}{[$\unit[]{ms}$]} & [\%] & \multicolumn{5}{c|}{[$\unit[]{ms}$]} & [\%] \\
      \hline
      \multirow{6}{*}{\begin{sideways}\emph{No Load}\end{sideways}}
      &\multirow{3}{*}{\begin{sideways}\emph{Night}\end{sideways}}
      & $\msg^{c1}_u$            & 1.040 &     1.442 &     19.567 &    25.523 &    72.141 &  0.00250
                                & 0.335 &     0.860 &     7.950 &    24.992 &    71.512 &  0.00250  \\
                              
      & & $\msg^{c165}_u$        & 0.302 &     1.160 &     18.416 &    20.234 &    20.378 &  0.0
                                & 0.079 &     0.052 &     0.255 &     0.409 &    29.220 &  0.0 \\
                             
      & & $\msg^{c1 \cup c165}_u$ & 0.209 &     0.099 &     1.394 &     3.864 &     8.249 &  0.0
                                & 0.077 &     0.050 &     0.182 &     0.319 &    29.220 &  0.0 \\
                                 
      \cline{2-15}
      & \multirow{3}{*}{\begin{sideways}\emph{Day}\end{sideways}}
      & $\msg^{c1}_u$            & 1.684 &     5.514 &   41.478 &   178.129 &   808.210 &  0.00111 
                                & 0.994 &     5.640 &   41.326 &   177.989 &   843.950 &  0.00111 \\
     & & $\msg^{c165}_u$         & 0.303 &     1.160 &   18.425 &    20.250 &    20.453 &  0.0
                                & 0.076 &     0.018 &    0.254 &     0.392 &     1.854 &  0.0 \\
     & & $\msg^{c1 \cup c165}_u$ & 0.213 &     0.211 &    2.455 &    10.123 &    19.943 &  0.0
                               & 0.074 &     0.013 &    0.176 &     \textbf{0.324} &     1.854 &  0.0 \\

     \hline
      \multirow{6}{*}{\begin{sideways}\emph{Low Load}\end{sideways}}
      & \multirow{3}{*}{\begin{sideways}\emph{Night}\end{sideways}}
      &  $\msg^{c1}_u$           &  1.049 &     1.469 &   19.632 &    24.980 &    85.837 &  0.000278 
                                &  0.342 &     0.853 &    8.304 &    22.519 &    85.194 &  0.000278 \\
      & & $\msg^{c165}_u$        &  0.437 &     1.250 &   18.959 &    21.087 &    22.955 &  0.0
                                &  0.220 &     0.312 &    2.486 &     6.087 &    14.920 &  0.0 \\
      & & $\msg^{c1 \cup c165}_u$ &  0.304 &     0.220 &   1.953 &     4.246 &    14.687 &  0.0
                                &  0.099 &     0.114 &   1.261 &     1.993 &     5.280 &  0.0 \\
      \cline{2-15}

      & \multirow{3}{*}{\begin{sideways}\emph{Day} \textbf{(b)}\end{sideways}}
      &  $\msg^{c1}_u$          &  1.236 &     1.750 &  20.207 &    29.519 &    86.122 &  0.00250
                               &  0.534 &     1.306 &  13.486 &    28.441 &    85.511 &  0.00278 \\
      & & $\msg^{c165}_u$       &  0.436 &     1.240 &  19.015 &    21.042 &    22.304 &  0.0
                               &  0.219 &     0.302 &   2.411 &     3.924 &    17.140 &  0.0 \\
      & & $\msg^{c1 \cup c165}_u$& 0.309 &     0.245 &   2.388 &     5.645 &    17.878 &  0.0
                               & 0.108 &     0.136 &   1.468 &     2.235 &     3.730 &  0.0 \\
      \hline

      \multirow{6}{*}{\begin{sideways}\emph{High Load}\end{sideways}}
      & \multirow{3}{*}{\begin{sideways}\emph{Night}\end{sideways}}
      & $\msg^{c1}_u$           & 1.125 &     1.566 &   19.823 &    27.197 &    46.916 &  0.000833
                               & 0.423 &     1.047 &   10.178 &    27.294 &    46.294 &  0.000833 \\
      & & $\msg^{c165}_u$       & 1.955 &     5.370 &   79.741 &   123.427 &   195.071 &  0.0 
                               & 1.746 &     5.236 &   79.555 &   123.332 &   194.985 &  0.0 \\
      & & $\msg^{c1 \cup c165}_u$& 0.606 &     0.523 &   5.322 &    13.218 &    \textbf{27.041} &  0.0
                               & 0.204 &     0.425 &   4.195 &     6.869 &    26.436 &  0.0 \\
       \cline{2-15}
      & \multirow{3}{*}{\begin{sideways}\emph{Day} \textbf{(c)}\end{sideways}}
      &  $\msg^{c1}_u$          & 1.285 &     4.521 &   20.306 &    99.607 &   \textbf{814.861} &  0.00250
                               & 0.584 &     4.367 &   13.372 &    99.030 &   812.695 &  0.00250 \\
      & & $\msg^{c165}_u$       & 1.746 &     4.554 &   65.367 &   119.014 &   160.995 &  0.0
                               & 1.539 &     4.395 &   65.105 &   118.924 &   160.891 &  0.0 \\
      & & $\msg^{c1 \cup c165}_u$& 0.624 &     0.555 &   5.696 &    11.893 &    22.976 &  0.0
                               & 0.236 &     0.477 &   4.572 &     \textbf{8.351} &    22.866 &  0.0 \\
      \hline

      \hline
    \end{tabular}
  \end{center}
\end{table*}

To better highlight the ACK frame collision issue, transmission requests on $\mathrm{I}$ were delayed by $\unit[10]{\mu s}$ with respect to $\mathrm{M}$.
The interaction between frames sent on the two channels is exemplified in the lower time diagram of Fig.~\ref{fig:aci}.
In particular, STA$_\mathrm{I}$ senses the channel busy and defers its transmission (situation is specular with respect to the previous experiment).
After one DIFS plus the backoff interval, computed from the end of the data frame on $\mathrm{M}$, a copy of the same packet is also sent on $\mathrm{I}$. 
If, depending on the random backoff selected by STA$_\mathrm{I}$, frame transmission on $I$ overlaps with the ACK frame on $\mathrm{M}$, reception of the latter by STA$_\mathrm{M}$ might sometimes fail because of proximity of antennas.

Results of experiments are reported in the lower part of Table~\ref{tab:ACI_results_new}. 
As expected, transmission latencies on $\mathrm{M}$ are unaffected. 
Conversely, ACI directly worsens $\mathrm{PLR^\prime}$ (but not $\mathrm{PLR}$). 
In fact, because of missed ACK reception, STA$_\mathrm{M}$ is wrongly led to believe that the packet was not correctly delivered to the AP, which in turn triggers a retransmission attempt.
However, since in the experiments we disabled automatic retransmission, a discrepancy appears concerning the transmission outcomes as seen on the originating and recipient sides.
For the $\overline{\ACI}$ set, the values of $\mathrm{PLR}$ and $\mathrm{PLR}^\prime$ are practically identical.
Instead, when frequencies of channels are close, for the $\ACI$ set there is a sensible increase of the number of dropped ACK frames ($\mathrm{PLR}^\prime=0.334\%$ vs. $\mathrm{PLR}=0.012\%$). 
Once again, this phenomenon completely disappears when channels are set to operate on different bands.

\subsubsection{Possible remedies to ACI}
In the above experiments, transmission times on $\mathrm{M}$ and $\mathrm{I}$ were displaced by $\unit[\pm 10]{\mu s}$ in order to always induce either frame delaying or ACK frame collision phenomena. 
In seamless redundancy implementations where packets are queued in the adapters of the originator at the same time, a mix of the above phenomena (plus, possibly, others) may take place, which worsens both latencies and losses. 
Variability of the related effects drastically increases due to retransmissions and rate adaptation techniques, which are typically enabled in real-world networks. 
As a consequence, to increase predictability and performance of seamless redundancy, ACI has to be prevented by setting redundant adapters on widely spaced channels or, even better, on different bands.
Alternatively, transmissions on redundant physical channels by the same RSTA must be adequately displaced in time, so as not to overlap.

\section{Performance of optimized redundant links}
\label{sec:red_results}

A final experimental campaign was performed after all the guidelines described in the previous sections were applied to the testbed.
Its aim is evaluating benefits that optimizations provide on seamless redundancy performance, by making communication paths practically independent.
In detail: 
\begin{enumerate}
	\item DTIM effects were prevented for multicast traffic,
	\item the NM service was disabled in the PC,
	\item two distinct APs were used as network infrastructure, and,
	\item channels were set to operate on different bands.
\end{enumerate}

Experiments were carried out both at \textit{Night} and by \textit{Day}, and lasted $10$ hours each.
Two packet streams $\msg^{c1}$ and $\msg^{c165}$ were sent on channels $1$ and $165$, respectively, and generation period was set to $T_c=\unit[100]{ms}$.
Since no STAs were found on channel $165$, besides those involved in the experiments (in the PC and the AP), load conditions on the two channels were pretty much unbalanced. 
This is the worst operating condition for links based on seamless redundancy, because the resulting communication quality tends to approach that of the best physical channel.
For this reason, besides the \textit{No Load} condition, we also defined the \textit{Low Load} and \textit{High Load} conditions, for which additional traffic was purposely injected on channel $165$. 
Both interfering loads are based on packet bursts, where packets with $\unit[1500]{B}$ payload size are generated every $\unit[400]{\mu s}$.
The number of packets in a burst and the gap between consecutive bursts are modeled according to exponential distributions, with mean values $300$ packets and $\unit[200]{ms}$, respectively. 
The difference between \textit{Low Load} and \textit{High Load} conditions lies in the number of interfering nodes, i.e., $1$ and $3$, respectively.

\subsection{Quality of Communication on the Redundant Link}
Both unicast and multicast traffics were taken into account. 
In the former case, we evaluated statistics on both the end-to-end latency $d$ measured on packet arrival at the PC Ethernet port (after traversing the AP) and the link latency $d^\prime$ obtained from ACK frame reception on wireless adapters. 
For multicast traffic, only results about end-to-end latency $d$ are available, because this kind of traffic is not confirmed.

\subsubsection{Unicast transmission}
Results are summarized in Table~\ref{tab:final_results_unicast}. 
As expected, tangible improvements were achieved by seamless redundancy in all experimental conditions and for every statistical index.
As an example, for the \emph{High Load--Day} condition, percentile $d_{p99.99}$ decreased by about one order of magnitude compared to single channels. 
Slightly lower advantages were achieved when individual channels are highly unbalanced, as in the \emph{No Load--Day} condition. 

Statistics on the end-to-end latency $d$ (left side of the table) mostly concern the point of view of applications. 
Think, e.g., to the communication quality perceived by a device (wireless sensor, AGV, drone, etc.) that communicates on air, by means of redundant APs, with a programmable logic controller (PLC) connected to the wired backbone.
In this case, results are biased by the characteristics of the specific APs used in the experiments.
For example, in our setup, every $\unit[10]{s}$ each AP causes an additional delay on frame forwarding between the wireless and wired segments, which may exceed $\unit[20]{ms}$ (see Section \ref{sec:NetworkComponents}).
The maximum latency recorded for single channels over all experiments was $\unit[814.861]{ms}$, which significantly decreases to $\unit[27.041]{ms}$ for the redundant link.

Instead, statistics on the link latency $d^\prime$ (right side of the table) do not depend in any way from APs.
Hence, results are more accurate when evaluating the redundant link alone. 
Surprisingly, the worst-case latency $d^{\prime}_{\operatorname{max}}$ occasionally exceeded $d_{\operatorname{max}}$ in the same experiment.
The explanation is that, the originator takes timestamps on ACK frame arrivals, but sometimes only the ACK frame, and not the related data frame, is lost, which means that a later ACK frame will be used for calculating $d^\prime$, thus making it possibly larger than $d$.

\subsubsection{Multicast transmission}
Experimental conditions are the same as for unicast packets. Results, reported in Table~\ref{tab:final_results_multicast}, show that the number of multicast packets lost on physical channels is much larger than in the unicast case. 
In fact, they are unconfirmed, and hence the retransmission mechanism cannot be exploited by the MAC layer to improve reliability.
While improvements provided by seamless redundancy on latency are not as good as for unicast transmissions, a dramatic reduction can be seen on the number of losses. 
For instance, in the \emph{High Load--Day} condition, only $\unit[1.658]{\%}$ of the packets were lost on the redundant link, while losses on individual channels amounted to $\unit[9.398]{\%}$ and $\unit[17.940]{\%}$, respectively.

Remarkably, the worst-case latency on the redundant link may sometimes coincide with physical channels. 
For instance, in the \emph{High Load--Night} condition, it is the same as channel~$1$, that is $\unit[133.312]{ms}$.
This situation was explicitly checked, and corresponds to a packet that suffered from the maximum delay on channel $1$ and went lost on channel $165$.

\subsection{Assessment of Independence between Paths}
To perform such an assessment we exploited the property that, when two events (either delays or losses) are statistically independent, their joint probability equals the product of individual probabilities.

\subsubsection{Unicast transmission}
In order to evaluate the effectiveness of guidelines to reduce dependence between channels, CCDFs can be considered \cite{2017-TII-PRP_REDUNDANCY}.
If paths are independent, the CCDF of latencies measured on the redundant link can be theoretically estimated from those of individual channels ($\mathrm{CCDF}^{c1}$ and $\mathrm{CCDF}^{c165}$) by means of Eq. (7) in \cite{2017-TII-PRP_REDUNDANCY}, i.e., $\widehat{\mathrm{CCDF}}^{c1 \cup c165} = \frac{1}
{1 - \mathrm{PLR}^{c1} \cdot \mathrm{PLR}^{c165}} \cdot \left[\mathrm{PLR}^{c1}\!\cdot\!(1\!-\!\mathrm{PLR}^{c165})\! \cdot\! \mathrm{CCDF}^{c165}\!+\!\mathrm{PLR}^{c165}\! \cdot\! (1\!-\!\mathrm{PLR}^{c1}) \right.\cdot$ \\
$\cdot \mathrm{CCDF}^{c1}\! + \left.\! (1\!-\!\mathrm{PLR}^{c165})\!\cdot\!(1\!-\!\mathrm{PLR}^{c1})\!\cdot\!\mathrm{CCDF}^{c1}\!\cdot\!\mathrm{CCDF}^{c165} \right]$,
where $\widehat{\mathrm{CCDF}}$ is the \emph{estimated} CCDF.

\begin{figure}[]
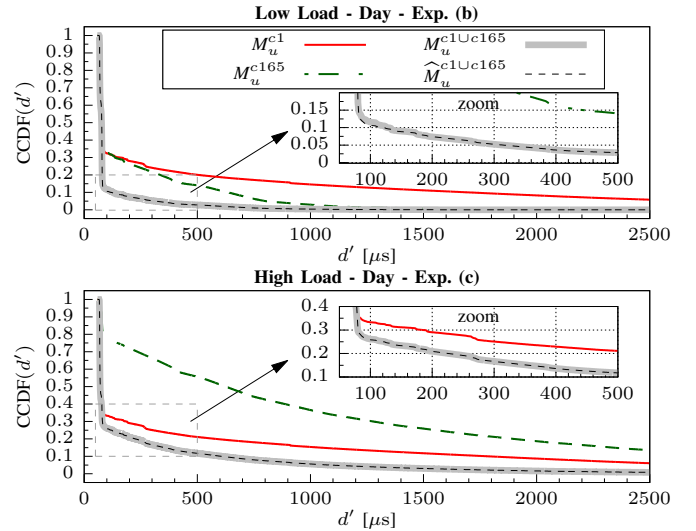

  \scriptsize
  \centering
  \include{FIG11-17-0869}
  \caption{CCDFs for experiments (b) and (c) in Table~\ref{tab:final_results_unicast} (latency derived from ACK frame arrival).}
  \label{fig:ccdffinal}
\end{figure}

As an example, for two specific experiments, identified in Table~\ref{tab:final_results_unicast} with labels (b) and (c), the measured CCDF of link latencies $d^{\prime}$ on the redundant link is reported in Fig.~\ref{fig:ccdffinal}, along with its theoretical estimate.
Since they practically coincide, paths can be considered independent.
In other words, the communication quality offered in practice by seamless redundancy is the same as what can be expected from theory, and can be modeled with excellent approximation using statistics acquired on individual Wi-Fi channels.

\subsubsection{Multicast transmission}
Quantity $\widehat{\mathrm{PLR}}$ in Table~\ref{tab:final_results_multicast} is the \emph{estimated} packet loss ratio, which in case channels are statistically independent can be obtained by multiplying their measured PLRs, i.e., $\widehat{\mathrm{PLR}}^{c1 \cup c165} = \mathrm{PLR}^{c1}\cdot \mathrm{PLR}^{c165}$ \cite{2017-TII-PRP_REDUNDANCY}.
As can be noted from results, the assumption of independence between paths reasonably holds for all the considered conditions.
In fact, estimated and measured PLR values were always similar, the only exception being the \emph{No Load--Night} condition, which is statistically unreliable because very few packets went lost.
Similarity between estimated and measured results, not reported here for space reasons, also holds for CCDFs of latencies for multicast packets.

\begin{table}[t]
  \caption{Results after optimization (multicast transmissions).}
  \label{tab:final_results_multicast}
  \scriptsize
  \begin{center}
    \tabcolsep=0.13cm
    \begin{tabular}{ccc||lllll|ll}
           &       &        & \multicolumn{6}{c}{Statistics derived from the PC Ethernet port} & \multirow{3}{*}{\begin{sideways}$\widehat{\mathrm{PLR}}$\end{sideways}} \\
      \multicolumn{2}{c}{Exp.} & Stream & $\bar{d}$ & $\sigma_{d}$ & $d_{p99.9}$ & $d_{p99.99}$  & $d_{\operatorname{max}}$ & $\mathrm{PLR}$ \\
        &  &  & \multicolumn{5}{c|}{[$\unit[]{ms}$]} & \multicolumn{2}{c}{[\%]} \\
      \hline
      \multirow{6}{*}{\begin{sideways}\emph{No Load}\end{sideways}}
      & \multirow{3}{*}{\begin{sideways}\emph{Night}\end{sideways}}

      & $\msg^{c1}_m$            & 1.382 & 1.553 & 19.770 & 27.163 & 47.791 & 6.131 & \multirow{3}{*}{\begin{sideways}\textbf{.000017}\end{sideways}}\\
      & & $\msg^{c165}_m$        & 0.862 & 1.162 & 18.932 & 20.749 & 86.359 & 0.000278 \\
      & & $\msg^{c1 \cup c165}_m$ & 0.772 & 0.328 &  3.105 & 17.181 & 86.359 & \textbf{0.000278} \\ 
      \cline{2-10}
      & \multirow{3}{*}{\begin{sideways}\emph{Day}\end{sideways}}

      & $\msg^{c1}_m$            & 1.873 & 3.816 & 30.004 & 84.203 & 796.201 & 9.731 & \multirow{3}{*}{\begin{sideways}\textbf{0.0}\end{sideways}}\\
      & & $\msg^{c165}_m$        & 0.862 & 1.163 & 18.920 & 20.723 & 54.372 & 0.0 \\
      & & $\msg^{c1 \cup c165}_m$ & 0.778 & 0.419 &  5.941 & 19.119 & 54.372 & \textbf{0.0} \\ 
      \hline
      \multirow{6}{*}{\begin{sideways}\emph{Low Load}\end{sideways}}

      & \multirow{3}{*}{\begin{sideways}\emph{Night}\end{sideways}}
      & $\msg^{c1}_m$            & 1.570 & 5.513 & 39.353 & 147.394 & 1145.965 & 5.574 & \multirow{3}{*}{\begin{sideways}\textbf{0.305}\end{sideways}}\\
      & & $\msg^{c165}_m$        & 0.982 & 1.209 & 19.189 & 21.321 & 22.548 & 5.477 \\
      & & $\msg^{c1 \cup c165}_m$ & 0.866 & 1.791 & 10.713 & 28.365 & 947.181 & \textbf{0.288} \\ 
      \cline{2-10}

      & \multirow{3}{*}{\begin{sideways}\emph{Day}\end{sideways}}
      & $\msg^{c1}_m$            & 1.594 & 2.280 & 20.988 & 60.583 & 329.424 & 7.748 & \multirow{3}{*}{\begin{sideways}\textbf{0.424}\end{sideways}} \\
      & & $\msg^{c165}_m$        & 0.981 & 1.216 & 19.351 & 21.247 &  22.912 & 5.469 \\
      & & $\msg^{c1 \cup c165}_m$ & 0.874 & 0.738 &  9.737 & 20.196 & 230.299 & \textbf{0.413} \\ 
      \hline
      \multirow{6}{*}{\begin{sideways}\emph{High Load}\end{sideways}}
      & \multirow{3}{*}{\begin{sideways}\emph{Night}\end{sideways}}

      & $\msg^{c1}_m$           & 1.365 & 1.609 & 19.887 & 29.653 & \textbf{133.312} & 6.483 & \multirow{3}{*}{\begin{sideways}\textbf{1.252}\end{sideways}} \\
      & & $\msg^{c165}_m$        & 1.622 & 1.494 & 20.311 & 22.592 & 62.383 & 19.318\\
      & & $\msg^{c1 \cup c165}_m$ & 1.073 & 0.890 & 13.810 & 21.210 & \textbf{133.312} & \textbf{1.252} \\ 
      \cline{2-10}

      & \multirow{3}{*}{\begin{sideways}\emph{Day}\end{sideways}}
      & $\msg^{c1}_m$            & 2.061 & 6.501 & 46.164 & 301.973 & 896.643 & \textbf{9.398} & \multirow{3}{*}{\begin{sideways}\textbf{1.685}\end{sideways}} \\
      & & $\msg^{c165}_m$        & 1.630 & 1.500 & 20.528 & 22.774 & 26.773 & \textbf{17.940} \\
      & & $\msg^{c1 \cup c165}_m$ & 1.260 & 2.424 & 20.011 & 58.843 & 579.104 & \textbf{1.658} \\ 
      \hline
    \end{tabular}
  \end{center}
\end{table}

\section{Conclusions}
Seamless redundancy according to PRP can be adopted to tangibly improve communication quality  in \mbox{Wi-Fi} networks.
However, to obtain the performance boost one may expect from theory, care has to be taken so that concurrent transmissions on physical paths of the same redundant link do not suffer from either joint or mutual interference.

In this paper, several phenomena have been considered that, if not properly addressed, may impair effectiveness of seamless redundancy.
They depend on aspects such as management activities performed in network equipment and end nodes, which may cause joint interference, antenna placement, which could lead to mutual interference on air between adjacent channels, and mechanisms like DTIM, which indirectly affect redundancy.
To practically analyze the effects of such phenomena on communication quality in redundant wireless network implementations, a thorough experimental campaign was carried out in a real setup, consisting of conventional Linux PCs and COTS Wi-Fi equipment.

Experiments enabled us to define suitable countermeasures.
From a practical viewpoint, the following guidelines should be followed when designing a networked control system communicating on redundant Wi-Fi links:
\begin{enumerate}[leftmargin=*]
	\item Mechanisms to support power saving in IEEE 802.11 battery-powered devices (e.g., DTIM) may severely interfere with timely delivery of multicast packets.
	To prevent this behavior, devices not belonging to the control system, like mobiles, should not be allowed to accidentally associate to the AP.
	This should be preferably obtained by enabling authentication and, possibly, encryption (e.g., WPA2).
	Although relevant, this aspect does not specifically concern redundant communications.
	
	\item Software modules typically exist in end-nodes, which may jointly affect transmissions on all paths of the redundant link.
	As far as possible, this has to be prevented.
	A relevant example is the network manager (NM), which is used to cope with device (re)association to the AP.
	Disabling NM is usually not tolerable, as reliability would be severely compromised.
	A viable solution is to modify the NM code (or to rewrite it from scratch), so that channel scanning is never carried out concurrently on redundant adapters.

	\item Joint interference caused by internal processing in network equipment (e.g., the AP) has to be prevented as well.
	Unfortunately, bringing modifications to the AP firmware is typically a quite complex task (and, sometimes, not possible at all).
In the case of commercial APs, interference can typically be mitigated by using a pair of distinct devices in place of a single simultaneous dual-band AP.

	\item RF modules may cause mutual interference between channels, when antennas are located too close to each other.
	In such conditions, adjacent channel interference (ACI) may occur, which prevents carrier sensing from operating correctly and affects communication on both paths at the same time.
	To prevent ACI, proper frequency selection is needed.
	While channels should be preferably chosen in distinct frequency bands ($\unit[2.4]{GHz}$ and $\unit[5]{GHz}$), widely-spaced channels in the $\unit[5]{GHz}$ band may also be used.
\end{enumerate}

Experimental evidences we collected confirm that the above guidelines make Wi-Fi channels statistically independent, to the point that communication quality on the redundant link matches the theoretical figures with very good approximation. 
Results, obtained from real testbeds configured in this way, fully support this claim.
Of course, several other minor aspects can be also considered, concerning both the protocol and its implementation.
They will be the subject of our future work.

\bibliographystyle{IEEEtran}
\bibliography{TII-17-0869.bib}

\begin{thebibliography}{10}
\providecommand{\url}[1]{#1}
\csname url@samestyle\endcsname
\providecommand{\newblock}{\relax}
\providecommand{\bibinfo}[2]{#2}
\providecommand{\BIBentrySTDinterwordspacing}{\spaceskip=0pt\relax}
\providecommand{\BIBentryALTinterwordstretchfactor}{4}
\providecommand{\BIBentryALTinterwordspacing}{\spaceskip=\fontdimen2\font plus
\BIBentryALTinterwordstretchfactor\fontdimen3\font minus
  \fontdimen4\font\relax}
\providecommand{\BIBforeignlanguage}[2]{{%
\expandafter\ifx\csname l@#1\endcsname\relax
\typeout{** WARNING: IEEEtran.bst: No hyphenation pattern has been}%
\typeout{** loaded for the language `#1'. Using the pattern for}%
\typeout{** the default language instead.}%
\else
\language=\csname l@#1\endcsname
\fi
#2}}
\providecommand{\BIBdecl}{\relax}
\BIBdecl

\bibitem{2016-std-80211}
``{IEEE Standard for Information technology--Telecommunications and information
  exchange between systems Local and metropolitan area networks--Specific
  requirements - Part 11: Wireless LAN Medium Access Control (MAC) and Physical
  Layer (PHY) Specifications},'' \emph{{IEEE Std 802.11-2016 (Rev. of IEEE Std
  802.11-2012)}}, pp. 1--3534, Dec 2016.

\bibitem{Willig_2008}
A.~Willig, ``Recent and emerging topics in wireless industrial communications:
  A selection,'' \emph{IEEE Trans. Ind. Informat.}, vol.~4, no.~2, pp.
  102--124, May 2008.

\bibitem{2011-IEM-Petersen}
S.~Petersen and S.~Carlsen, ``{WirelessHART Versus ISA100.11a: The Format War
  Hits the Factory Floor},'' \emph{IEEE Industrial Electronics Magazine},
  vol.~5, no.~4, pp. 23--34, Dec 2011.

\bibitem{2007-TII-vas}
R.~Moraes, F.~Vasques, P.~Portugal, and J.~Fonseca, ``{VTP-CSMA: A Virtual
  Token Passing Approach for Real-Time Communication in IEEE 802.11 Wireless
  Networks},'' \emph{IEEE Trans. Ind. Informat.}, vol.~3, no.~3, pp. 215--224,
  2007.

\bibitem{2011-ISPCS-TDMA-flexWARE}
H.~Trsek and J.~Jasperneite, ``{An isochronous medium access for real-time
  wireless communications in industrial automation systems - A use case for
  wireless clock synchronization},'' in \emph{IEEE Int. Symp. on Precision
  Clock Synchronization for Measurement Control and Communication (ISPCS)},
  Sept 2011, pp. 81--86.

\bibitem{Gamba_et_al_2010}
G.~Gamba, F.~Tramarin, and A.~Willig, ``{Retransmission Strategies for Cyclic
  Polling Over Wireless Channels in the Presence of Interference},''
  \emph{{IEEE Trans. Ind. Informat.}}, vol.~6, no.~3, pp. 405--415, Aug 2010.

\bibitem{2017-TII-EDF}
L.~Seno, G.~Cena, S.~Scanzio, A.~Valenzano, and C.~Zunino, ``{Enhancing
  Communication Determinism in Wi-Fi Networks for Soft Real-Time Industrial
  Applications},'' \emph{IEEE Trans. Ind. Informat.}, vol.~13, no.~2, pp.
  866--876, April 2017.

\bibitem{2016-TII-Tian}
G.~Tian, S.~Camtepe, and Y.~C. Tian, ``{A Deadline-Constrained 802.11 MAC
  Protocol With QoS Differentiation for Soft Real-Time Control},'' \emph{IEEE
  Trans. Ind. Informat.}, vol.~12, no.~2, pp. {544--554}, April 2016.

\bibitem{2010-TII-802.11e}
G.~Cena, L.~Seno, A.~Valenzano, and C.~Zunino, ``{On the Performance of IEEE
  802.11e Wireless Infrastructures for Soft-Real-Time Industrial
  Applications},'' \emph{IEEE Trans. Ind. Informat.}, vol.~6, no.~3, pp.
  425--437, 2010.

\bibitem{2008-IEM-Hybrid}
G.~Cena, A.~Valenzano, and S.~Vitturi, ``Hybrid wired/wireless networks for
  real-time communications,'' \emph{IEEE Industrial Electronics Magazine},
  vol.~2, no.~1, pp. 8--20, March 2008.

\bibitem{2016-TIM-EWMA}
T.~D. Chung, R.~B. Ibrahim, V.~S. Asirvadam, N.~B. Saad, and S.~M. Hassan,
  ``{Adopting EWMA Filter on a Fast Sampling Wired Link Contention in
  WirelessHART Control System},'' \emph{IEEE Trans. Instrum. Meas.}, vol.~65,
  no.~4, pp. 836--845, April 2016.

\bibitem{2016-SGComm}
M.~Mohiuddin, M.~Popovic, A.~Giannakopoulos, and J.~Y.~L. Boudec,
  ``{Experimental validation of the usability of Wi-Fi over redundant paths for
  streaming phasor data},'' in \emph{Int. Conference on Smart Grid
  Communications (SmartGridComm)}, Nov 2016, pp. 533--538.

\bibitem{2017-TII-PRP_REDUNDANCY}
G.~Cena, S.~Scanzio, and A.~Valenzano, ``{Experimental Evaluation of Seamless
  Redundancy Applied to Industrial Wi-Fi Networks},'' \emph{IEEE Trans. Ind.
  Informat.}, vol.~13, no.~2, pp. 856--865, April 2017.

\bibitem{2016-ETFA-Guidelines}
------, ``{Design guidelines to improve reliability of seamless redundancy in
  Wi-Fi networks},'' in \emph{IEEE Int. Conf. on Emerging Technologies and
  Factory Automation (ETFA)}, Sept 2016, pp. 1--8.

\bibitem{2012-std-PRP}
{IEC}, ``{Industrial communication networks - High availability automation
  networks - Part 3: Parallel Redundancy Protocol (PRP) and High-availability
  Seamless Redundancy (HSR)},'' \emph{IEC Std 62439-3 ed 2.0}, pp. 1--177,
  2010.

\bibitem{2015-TIM-PRP_wired_wireless}
P.~Castello, P.~Ferrari, A.~Flammini, C.~Muscas, P.~A. Pegoraro, and
  S.~Rinaldi, ``{A Distributed PMU for Electrical Substations With Wireless
  Redundant Process Bus},'' \emph{IEEE Trans. Instrum. Meas.}, vol.~64, no.~5,
  pp. 1149--1157, May 2015.

\bibitem{2016-TII-iPRP}
M.~Popovic, M.~Mohiuddin, D.~C. Tomozei, and J.~Y.~L. Boudec, ``{iPRP---The
  Parallel Redundancy Protocol for IP Networks: Protocol Design and
  Operation},'' \emph{IEEE Trans. Ind. Informat.}, vol.~12, no.~5, pp.
  1842--1854, Oct 2016.

\bibitem{2017-TII-survey_sync}
A.~Mahmood, R.~Exel, H.~Trsek, and T.~Sauter, ``{Clock Synchronization Over
  IEEE 802.11; A Survey of Methodologies and Protocols},'' \emph{IEEE Trans.
  Ind. Informat.}, vol.~13, no.~2, pp. 907--922, April 2017.

\bibitem{2017-EUROCON-redundancy_railway}
M.~M. Awad, H.~H. Halawa, M.~Rentschler, R.~M. Daoud, and H.~H. Amer, ``{Novel
  system architecture for railway wireless communications},'' in \emph{IEEE
  Int. Conference on Smart Technologies (EUROCON)}, July 2017, pp. 877--882.

\bibitem{2014-WFCS-Mifdaoui}
D.~K. Dang, A.~Mifdaoui, and T.~Gayraud, ``{Design and analysis of UWB-based
  network for reliable and timely communications in safety-critical
  avionics},'' in \emph{IEEE Workshop on Factory Communication Systems (WFCS)},
  May 2014, pp. 1--10.

\bibitem{2012-WFCS-WoP1}
M.~Rentschler and P.~Laukemann, ``{Towards a reliable parallel redundant WLAN
  black channel},'' in \emph{IEEE Int. Workshop on Factory Communication
  Systems (WFCS)}, 2012, pp. 255--264.

\bibitem{2013-ETFA-Rentschler}
M.~Rentschler, O.~A. Mady, M.~T. Kassis, H.~H. Halawa, T.~K. Refaat, R.~M.
  Daoud, H.~H. Amer, and H.~M. ElSayed, ``{Simulation of parallel redundant
  WLAN with OPNET},'' in \emph{IEEE Conference on Emerging Technologies Factory
  Automation (ETFA)}, Sept 2013, pp. 1--8.

\bibitem{2016-tii-WiRed}
G.~Cena, S.~Scanzio, and A.~Valenzano, ``{Seamless Link-Level Redundancy to
  Improve Reliability of Industrial Wi-Fi Networks},'' \emph{IEEE Trans. Ind.
  Informat.}, vol.~12, no.~2, pp. 608--620, April 2016.

\bibitem{2014-WFCS-WiRed}
G.~Cena, S.~Scanzio, A.~Valenzano, and C.~Zunino, ``{An enhanced MAC to
  increase reliability in redundant Wi-Fi networks},'' in \emph{IEEE Int.
  Workshop on Factory Communication Systems (WFCS)}, May 2014, pp. 1--10.

\bibitem{2014-ETFA-DDD}
------, ``{Dynamic duplicate deferral techniques for redundant Wi-Fi
  networks},'' in \emph{IEEE Int. Conf. on Emerging Technology and Factory
  Automation (ETFA)}, Sept 2014, pp. 1--8.

\bibitem{2017-WFCS-bandwidth}
G.~Cena, S.~Scanzio, and A.~Valenzano, ``{Duplication avoidance mechanisms to
  reduce bandwidth usage in redundant Wi-Fi networks},'' in \emph{IEEE Int.
  Workshop on Factory Communication Systems (WFCS)}, May 2017, pp. 1--10.

\bibitem{2017-WFCS-SDMAC}
------, ``{A software-defined MAC architecture for Wi-Fi operating in user
  space on conventional PCs},'' in \emph{IEEE Int. Workshop on Factory
  Communication Systems (WFCS)}, May 2017, pp. 1--10.

\bibitem{2015-std-etsien300328}
``{Electromagnetic compatibility and Radio spectrum Matters (ERM); Wideband
  transmission systems; Data transmission equipment operating in the 2,4 GHz
  ISM band and using wide band modulation techniques; Harmonized EN covering
  the essential requirements of article 3.2 of the R\&TTE Directive},''
  \emph{ETSI EN 300 328 V1.9.1 Std}, pp. 1--91, Feb. 2015.

\bibitem{2014-ETFA-trama}
F.~Tramarin, S.~Vitturi, M.~Luvisotto, and R.~Parrozzani, ``{Performance
  assessment of an IEEE 802.11-based protocol for real-time communication in
  agriculture},'' in \emph{IEEE Int. Conf. on Emerging Technology and Factory
  Automation (ETFA)}, Sept 2014, pp. 1--6.

\bibitem{2008-IWCMCC-ACI}
J.~Nachtigall, A.~Zubow, and J.~P. Redlich, ``{The Impact of Adjacent Channel
  Interference in Multi-Radio Systems using IEEE 802.11},'' in \emph{Int.
  Wireless Communications and Mobile Computing Conference}, Aug 2008, pp.
  874--881.

\bibitem{2007-TII-PDF}
G.~Cena, I.~C. Bertolotti, A.~Valenzano, and C.~Zunino, ``{Evaluation of
  Response Times in Industrial WLANs},'' \emph{{IEEE Trans. Ind. Informat.}},
  vol.~3, no.~3, pp. 191--201, Aug 2007.

\end{thebibliography}

\begin{IEEEbiography}%
[{\includegraphics[width=1in,height=1.25in,clip,keepaspectratio]
{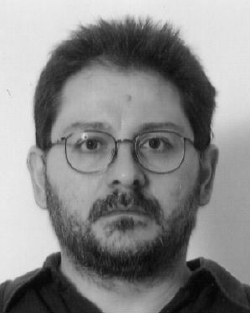}}]{Gianluca Cena} (SM'09) received the Laurea degree in electronic engineering and the Ph.D. degree in information and system engineering from the Politecnico di Torino, Italy, in 1991 and 1996, respectively.

Since 2005 he has been Director of Research with the Institute of Electronics, Computer and Telecommunication Engineering, National Research Council of Italy (CNR--IEIIT), Turin. His research interests include wired and wireless industrial communication systems, real-time protocols, and automotive networks. In these areas he has coauthored about 130 technical papers, three of which awarded as Best Papers of the 2004, 2010, and 2017 editions of the IEEE Workshops on Factory Communication Systems, and one international patent. 

Dr. Cena served as a Program Co-Chairman for the 2006 and 2008 editions of the IEEE International Workshop on Factory Communication Systems, and as a Track Co-Chairman in six editions of the IEEE International Conference on Emerging Technologies and Factory Automation. Since 2009 he has been an Associate Editor of the IEEE Transactions on Industrial Informatics.
\end{IEEEbiography}

\begin{IEEEbiography}%
[{\includegraphics[width=1in,height=1.25in,clip,keepaspectratio]
{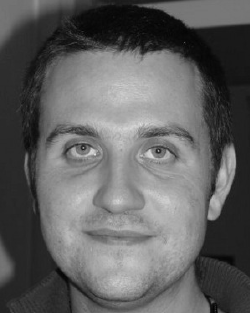}}]{Stefano Scanzio} (S'06-M'12) received the Laurea and Ph.D. degrees in Computer Science from Politecnico di Torino, Torino, Italy, in 2004 and 2008, respectively.

He was with the Department of Computer Engineering, Politecnico di Torino, from 2004 to 2009, where he was involved in research on speech recognition and, in particular, he has been active in classification methods and algorithms. Since 2009, he has been with the National Research Council of Italy (CNR), where he is a tenured Researcher with the Institute of Electronics, Computer and Telecommunication Engineering (IEIIT), Turin.

Dr. Scanzio teaches several courses on Computer Science at Politecnico di Torino. He has authored and co-authored of more than 50 papers in international journals and conferences, most of them in the area of industrial communication systems, real-time networks, wireless networks and clock synchronization protocols. He received the Best Paper Awards for the papers he presented at the 8th and 13th IEEE Workshops on Factory Communication Systems (WFCS 2010 and WFCS 2017).
\end{IEEEbiography}

\begin{IEEEbiography}%
[{\includegraphics[width=1in,height=1.25in,clip,keepaspectratio]
{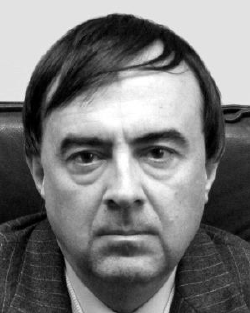}}]{Adriano Valenzano} (SM'09) received the Laurea degree in electronic engineering from Politecnico di Torino, Torino, Italy, in 1980.

He is Director of Research with the National Research Council of Italy (CNR). He is currently with the Institute of Electronics, Computer and Telecommunication Engineering (IEIIT), Torino, Italy, where he is responsible for research concerning distributed computer systems, local area networks, and communication protocols. He has coauthored approximately 200 refereed journal and conference papers in the area of computer engineering.

Dr. Valenzano is the recipient of the 2013 IEEE IES and ABB Lifetime Contribution to Factory Automation Award. He also received, as a coauthor, the Best Paper Award presented at the 5th, 8th and 13th IEEE Workshops on Factory Communication Systems (WFCS 2004, WFCS 2010 and WFCS 2017). He has served as a technical referee for several international journals and conferences, also taking part in the program committees of international events of primary importance. Since 2007, he has been serving as an Associate Editor for the IEEE Transactions on Industrial Informatics.
\end{IEEEbiography}

\vfill

\end{document}